\documentclass[prx,twocolumn,english,superscriptaddress,floatfix,longbibliography]{revtex4-2}

\usepackage{amsmath}
\usepackage{amssymb}
\usepackage{svg}
\usepackage[separate-uncertainty=true]{siunitx}
\definecolor{bluepoli}{RGB}{0,36,179}
\definecolor{redpoli}{RGB}{204,0,51}
\definecolor{greenpoli}{RGB}{45,137,0}
\definecolor{purplepoli}{RGB}{153,102,204}
\definecolor{azzurropoli}{RGB}{51,53,204}
\definecolor{orangepoli}{RGB}{255,124,17}
\usepackage{hyperref}
\usepackage{physics}
\usepackage{upgreek} 
\hypersetup{%
	colorlinks,
	linkcolor={bluepoli}, %
	citecolor={redpoli}, %
	urlcolor={redpoli} %
}

\def\QDevaffil{Center for Quantum Devices, Niels Bohr Institute, University of Copenhagen, 2100 Copenhagen, Denmark}
\def\RLEaffil{Research Laboratory of Electronics, Massachusetts Institute of Technology, Cambridge, MA 02139, USA}
\def\NTNUaffil{Center for Quantum Spintronics, Department of Physics,
	Norwegian University of Science and Technology, NO-7491 Trondheim, Norway}
\def\Physaffil{Department of Physics, Massachusetts Institute of Technology, Cambridge, MA 02139, USA}
\def\EECSaffil{Department of Electrical Engineering and Computer Science, Massachusetts Institute of Technology, Cambridge, MA 02139, USA}
\def\LLaffil{Lincoln Laboratory, Massachusetts Institute of Technology, Lexington, MA 02421, USA}

\begin{document}
	
	
	\title{Efficient Qubit Calibration by Binary-Search Hamiltonian Tracking}
	
	\author{Fabrizio~Berritta}
	\affiliation{\RLEaffil}	
	\affiliation{\QDevaffil}	
	\author{Jacob~Benestad}
	\affiliation{\NTNUaffil}
	\author{Lukas~Pahl}
	\affiliation{\RLEaffil}
	\affiliation{\EECSaffil}
	\author{Melvin Mathews}
	\affiliation{\RLEaffil}
	\affiliation{Department of Information Technology and Electrical Engineering, ETH Z\"urich, 8093 Z\"urich, Switzerland}
	\author{Jan~A.~Krzywda}
	\affiliation{$\langle a Q a^L \rangle$ Applied Quantum Algorithms --- Lorentz Institute for Theoretical Physics \& Leiden Institute of Advanced Computer Science, Universiteit Leiden, The Netherlands}
	\author{R\'eouven~Assouly}
	\affiliation{\RLEaffil}
	\author{Youngkyu~Sung}
	\affiliation{\RLEaffil}
	\affiliation{\EECSaffil}
	\author{David~K.~Kim}
	\affiliation{\LLaffil}
	\author{Bethany~M.~Niedzielski}
	\affiliation{\LLaffil}
	\author{Kyle~Serniak}
	\affiliation{\RLEaffil}
	\affiliation{\LLaffil}
	\author{Mollie~E.~Schwartz}
	\affiliation{\LLaffil}
	\author{Jonilyn~L.~Yoder}
	\affiliation{\LLaffil}
	\author{Anasua~Chatterjee}
	\affiliation{\QDevaffil}	
	\affiliation{
		QuTech and Kavli Institute of Nanoscience, Delft University of Technology, Delft, The Netherlands}	
	\author{Jeffrey~A.~Grover}
	\affiliation{\RLEaffil}
	\author{Jeroen~Danon}
	\affiliation{\NTNUaffil}
	\author{William~D.~Oliver}
	\affiliation{\RLEaffil}
	\affiliation{\EECSaffil}
	\affiliation{\Physaffil}
	\author{Ferdinand~Kuemmeth}
	\email{ferdinand.kuemmeth@ur.de}
	\affiliation{\QDevaffil}	
	\affiliation{Institute of Experimental and Applied Physics, University of Regensburg, 93040 Regensburg, Germany}
	\affiliation{QDevil, Quantum Machines, 2750 Ballerup, Denmark}


	
	\date{August 26, 2025}
	
	\begin{abstract}
       We present and experimentally implement a real-time protocol for calibrating the frequency of a resonantly driven qubit, achieving exponential scaling in calibration precision with the number of measurements, up to the limit imposed by decoherence. The real-time processing capabilities of a classical controller dynamically generate adaptive probing sequences for qubit-frequency estimation. Each probing evolution time and drive frequency are calculated to divide the prior probability distribution into two branches, following a locally optimal strategy that mimics a conventional binary search. The scheme does not require repeated measurements at the same setting, as it accounts for state preparation and measurement errors. Its use of a parametrized probability distribution favors numerical accuracy and computational speed. 
	We show the algorithm's efficacy by stabilizing a flux-tunable transmon qubit, leading to improved coherence and gate fidelity.
        As benchmarked by gate set tomography, the FPGA-powered control electronics partially mitigates non-Markovian noise, which is detrimental to quantum error correction. The mitigation is achieved by dynamically updating and feeding forward the qubit frequency.
        Our protocol highlights the importance of feedback in improving the calibration and stability of qubits subject to drift and can be readily applied to other qubit platforms.
		
	\end{abstract}
	\maketitle
	
	\section{Introduction}
	\begin{figure}
		\centering
		\includegraphics{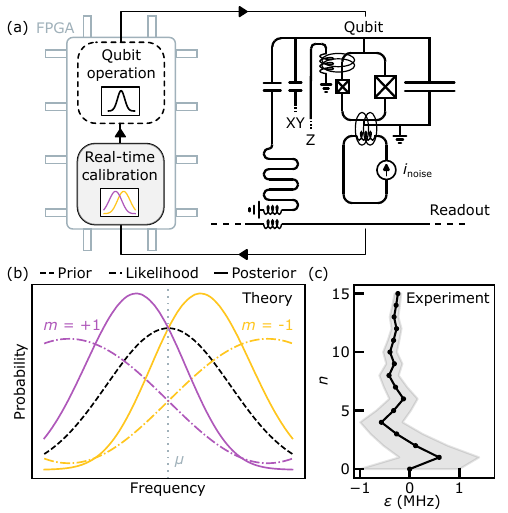}
		\caption{\textbf{Qubit implementation and frequency binary search}. 
			\textbf{(a)} Experiment schedule, alternating between periods of quantum information processing (dashed box), and periods for efficiently calibrating the qubit frequency (gray box). The transmon (Qubit) is controlled by microwave pulses (XY) that are based on previous qubit measurements (Readout). 
			\textbf{(b)} Evolution of the probability distribution ${\cal P}(\varepsilon)$ during the frequency binary search algorithm. For each Ramsey probing cycle, the probing time and frequency detuning are chosen such that the two possible likelihood functions (dash--dotted lines) multiplied with the prior distribution (dashed) yield posterior distributions (solid) that are shifted left or right, while the standard deviation is reduced (see main text).
			\textbf{(c)} The controller updates the probability distribution after each measurement. The dots show the resulting expectation value $\langle \varepsilon \rangle$ and the shaded area marks the 68\% credible interval.
		}
		\label{fig:fig1_FBS}
	\end{figure}
	Quantum processing units (QPUs) with tens of qubits are becoming increasingly common~\cite{Ichikawa2024}. Efforts to reduce error rates have resulted in the first demonstrations of quantum error correction~\cite{Campbell2024, Acharya2024}, though these advances come with costly calibration overhead as the number of qubits increases. Temporal instability in QPUs arises from numerous stochastic noise channels~\cite{Ball2016}, and the lowest-performing outlier qubits often limit overall performance~\cite{mohseni2024}.
	Tackling time-dependent fluctuations and outlier qubits in large QPUs requires scalable and efficient calibration methods to ensure stable, fault-tolerant operation while minimizing calibration downtime. 
	We introduce here a protocol that features exponential scaling of the calibration precision versus the number of measurements until it is limited by decoherence. 
	This scaling is the result of a locally optimal strategy that maximizes the expected precision for the next measurement.
	
	Online Hamiltonian learning~\cite{Gebhart2023} offers a promising strategy for addressing drifts in stochastic qubit parameters through real-time estimation~\cite{Gebhart2023, reuer2023realizing, Arshad2024, Berritta2024a,dumoulin2024silicon, vora2024ml, Berritta2024b, Park2025}, facilitated by modern field-programmable gate array (FPGA) hardware advancements. Although various estimation methods~\cite{Kimmel2015,mcmichael2022, Hurant2024, de2024time} have been proposed to enhance calibration efficiency, no experimental implementation has yet achieved real-time adaptive estimation in a resonantly driven qubit with a parametrized probability distribution. This work fills the gap by the experimental demonstration of a real-time Bayesian estimation approach applied to a superconducting qubit.

	To demonstrate the adaptive and efficient calibration protocol, we employ a flux-tunable transmon qubit~\cite{Krantz2019, Blais2021}. 
	Transmons are largely insensitive to charge noise by design. The tunable qubit frequency is controlled by a magnetic field and it is susceptible to random flux variations, known as flux noise.
	Typically, flux-tunable transmons are thus operated at bias points (sweet spots) that are first-order insensitive to small flux changes. However, operating them away from these bias points may be necessary to reduce frequency crowding in large qubit arrays~\cite{hertzberg2021laser, berke2022transmon} or minimize coupling to two-level fluctuators. To maintain high-fidelity operation despite the increased sensitivity to flux noise, one may use real-time frequency estimation and stabilization~\cite{Vepsaelaeinen2022}. 
	
	The so-called frequentist approach is the most common method for estimating fluctuations in Hamiltonian parameters of superconducting qubits. Frequentist estimation methods infer an observable from the observed frequencies of measurement outcomes in repeated experiments and they do not yield an exponential scaling with the number of measurements. For instance, in Ref.~\cite{Vepsaelaeinen2022} the qubit frequency fluctuation is computed in real time by repeated Ramsey interferometry cycles. In the probing Ramsey cycle, the Bloch vector is initialized on the equator of the Bloch sphere, evolves under its Hamiltonian, and is projected back onto the $\hat{z}$-axis after a fixed (\emph{non-adaptive}) evolution time. The measurement outcome is averaged over many probing cycles. The fraction of times the system is measured in the excited state, for instance, is linearly mapped to the qubit frequency fluctuation, as further detailed in the next section. The drawbacks of such frequentist methods are that (i) they are not optimally efficient, (ii) there is a trade-off between frequency sensitivity and probing range, and (iii) the qubit must already be approximately calibrated for the stabilization to work.
	In contrast, a Bayesian estimation approach assigns probabilities to hypotheses and naturally incorporates real-time optimization techniques~\cite{Gebhart2023, Arshad2024, Berritta2024a, Berritta2024b, Park2025}. At any given time, the current knowledge of the parameters can be used to \emph{adaptively} select the optimal experimental settings for the next probing cycle. While such techniques have been widely adopted in nitrogen-vacancy centers and semiconductor spin qubits over the past decade, they have not been similarly applied to superconducting qubits.
	
	In this work, we employ an adaptive Bayesian estimation scheme to calibrate a transmon qubit, which circumvents the limitations of the frequentist approach mentioned above.
	Our protocol involves programming a commercial controller with an integrated FPGA to generate dynamic microwave pulses, enabling real-time qubit frequency calibration through a binary search algorithm inspired by Ref.~\cite{Ferrie2013}. We use the real-time capabilities of the controller (on the qubit coherence timescale) to select optimal parameters within an adaptive estimation cycle. Specifically, the controller dynamically adjusts the drive frequency and the evolution time of each probing cycle based on previous measurement results. We demonstrate the algorithm's efficacy through improved effective qubit coherence $T_2^*$ and fidelity.
	
	While the majority of theoretical research concentrates on Markovian noise~\footnote{The commonly used term \emph{non-Markovian noise} refers to the non-unitary evolution of a quantum system that cannot be modeled by the Markovian master equation. Such non-Markovian decoherence arises from the environment's finite memory, which introduces correlations between consecutive measurement outcomes.}, non-Markovian noise sources such as $1/f$ flux noise, which exhibit memory effects, are common in solid-state qubit platforms~\cite{Park2025}. 
	They introduce significant overhead for error mitigation~\cite{hakoshima2021relationship, kam2024detrimental}, their dynamical decoupling~\cite{Viola1999, Szankowski2017} is not universally effective and may not align with specific experimental goals.
	In order to realize fault-tolerant quantum computing with error-corrected solid-state qubits, non-Markovian noise likely needs to be reduced or eliminated~ \cite{pataki2024coherent}. We show how our real-time frequency tracking protocol reduces such noise as validated by gate set tomography~\cite{Nielsen2021}, whereas Ref.~\cite{Vepsaelaeinen2022} relied solely on randomized benchmarking~\cite{Knill2008}. While randomized benchmarking blends the non-Markovian noise with other error types by introducing random gate sequences, gate set tomography assumes Markovian noise and can therefore detect non-Markovianity in the system through model violations.  
	Our scheme enables intermittent and efficient calibration of qubit frequencies, making it ideal for stable quantum circuit execution in the presence of drift.
	
	\section{Method}
	
	\subsection{Device}

	We use a $2\times2$ superconducting qubit array operated at the mixing-chamber stage (below $\SI{30}{\milli\kelvin}$) of a dilution refrigerator. The array design is based on Ref.~\cite{Sung2021}, and we calibrate one of its flux-tunable transmon qubits. A commercial controller (Quantum Machines OPX+ and Octave) applies high-frequency waveforms to the lines for qubit control and single-shot readout. A Yokogawa GS200 provides the DC flux bias through a bias-tee at room temperature (see the Supplemental Material~\cite{supplementary} for details on the experimental setup). The transmon comprises a DC SQUID with asymmetric junctions, and on-chip Z and XY control lines [see Fig.~\ref{fig:fig1_FBS}(a)]. The qubit is dispersively coupled to a coplanar waveguide resonator for state readout~\cite{Krantz2019, Blais2021}.\\
	
	The real-time calibration we develop here can be embedded in an operation loop as sketched in Fig.~\ref{fig:fig1_FBS}(a). The calibration protocol is used to estimate the qubit frequency, after which the resulting estimate is used to adjust the frequency of the drive accordingly, leading to a decreased sensitivity to flux noise during qubit operation.
	
	Our goal is to compensate for qubit frequency fluctuations particularly when it is operated away from the sweet spot. Thus we tune the transmon where the effect of flux noise is the strongest, at $\Phi_{\text{ext}}=\Phi_0/4$, where $\Phi_0 = h/(2e)$ is the superconducting magnetic flux quantum and $\Phi_{\text{ext}}$ is the external magnetic flux applied via a mutual coupling to the Z line. Our qubit nominally has a transition frequency $f_{\text{q}} \approx \SI{3.78}{\giga\hertz}$
	and in the rotating frame its Hamiltonian is
	\begin{equation}
		\frac{\mathcal{H}(t)}{h}=-\dfrac{\Updelta f - \varepsilon(t)}{2}\sigma_z,
	\end{equation}
	where $\sigma_{z}$ is the Pauli $z$-matrix, $\Updelta f=f_{\text{d}}-f_{\text{q}}$ is the frequency detuning between $f_{\text{q}}$ and the chosen rotating frame frequency $f_{\text{d}}$, and $\varepsilon(t)$ represents the time-dependent shift of the qubit frequency due to flux noise.
	
	\subsection{Frequency binary search by Bayesian estimation}
	\begin{figure*}
		\centering
		\includegraphics{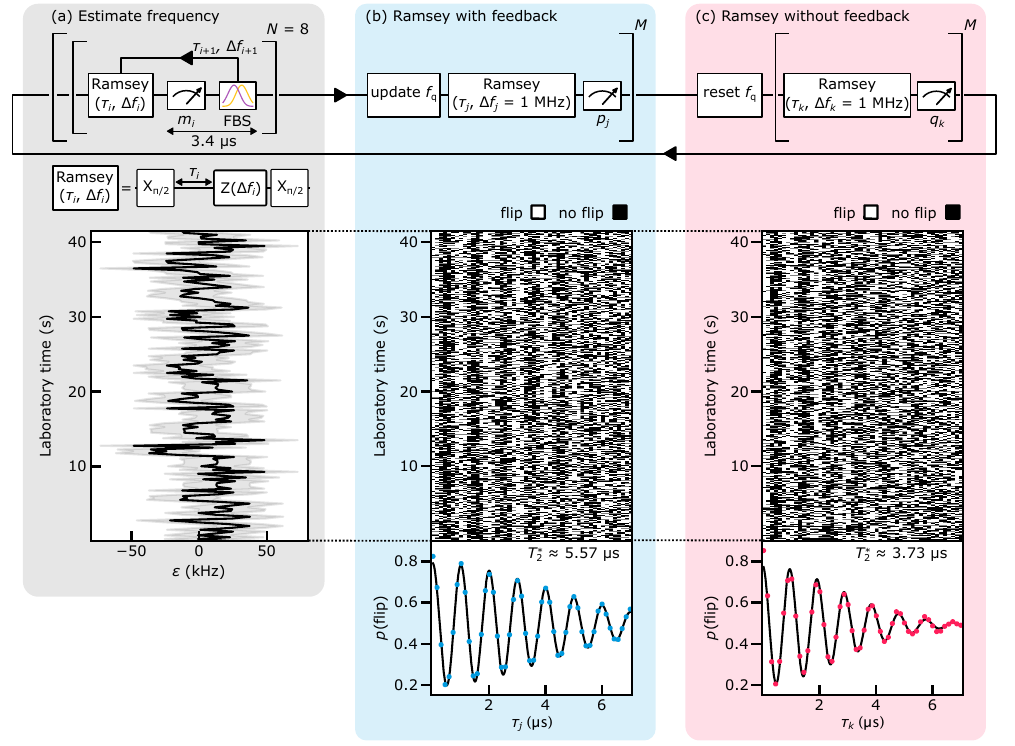}
		\caption{\textbf{Frequency binary search validation by suppressed dephasing of a qubit in a feedback-controlled rotating frame}.
			(upper diagram) One loop (solid arrows) represents one repetition of the protocol. 
			\textbf{(a)} For each estimate of the frequency shift $\varepsilon$, the controller uses $N=8$ Ramsey probing cycles with adaptive probe times $\tau_i$ and detuned frequencies $\Updelta f_i$, computed in real time. Bottom panel: $\langle \varepsilon \rangle$ (solid line) and 68\% credible interval (shaded area) of the final probability distribution ${\cal P}_8(\varepsilon)$ of all estimates performed during the $\approx \SI{41.5}{\second}$ of the experiment. 
			\textbf{(b)} After each estimation, we task the controller to perform a Ramsey cycle with evolution time $\tau_j$, while setting $\Updelta f_j = \SI{1}{\mega\hertz}$ by adjusting $f_{\rm d}$ using the latest estimate $\langle \varepsilon\rangle$. These interleaved FBS estimations and Ramsey cycles are repeated $M=50$ times with evenly spaced $\tau_j$. 
			Single-shot measurement outcomes $p_j$ are plotted in the middle panel as white/black pixels, and the fraction of flipped outcomes in each column is shown as a blue dot in the lower panel.
			\textbf{(c)} Subsequently the controller resets all frequencies to the offline-calibrated values, i.e., assumes $\langle \varepsilon \rangle = 0$, and performs again $M=50$ evenly spaced Ramsey cycles of set $\Updelta f_k = \SI{1}{\mega\hertz}$. Single-shot measurement outcomes $q_k$ are plotted as white/black pixels in the middle panel, and the fraction of flipped outcomes in each column is shown as a red dot in the lower panel. Comparing (b) to (c), the coherence time improves by $\approx49\%$ with feedback.
		}
		\label{fig:fig2_FBS}
	\end{figure*}
	The frequency binary search (FBS) algorithm involves a Ramsey probing cycle where the qubit is first prepared in a superposition state $\ket{\psi} = \left( \ket{0} - i \ket{1} \right)/\sqrt{2}$ using a X$_{\pi/2}$ rotation around the $\hat{x}$-axis of the Bloch sphere. The state preparation is followed by a period of free evolution for a duration $\tau$, during which the qubit state acquires a phase $
	\phi = 2\pi \int_0^\tau \delta_\text{q}(t') \, \mathrm{d}t'$, where $\delta_\text{q}(t) = \Updelta f - \varepsilon(t)$. A second X$_{\pi/2} $ pulse is then applied, and the state of the qubit is measured using dispersive readout. By the quasistatic noise approximation, we assume $\varepsilon(t)$ to be constant on the scale of a few probing cycles, resulting in $\phi = 2\pi[ \Updelta f - \varepsilon(t)]\tau$. In the following, we drop the time dependence of $\varepsilon(t)$ for ease of notation.
	
	We assume that the probability of measuring an outcome $m\in \{-1,1\}$ corresponding to the states $|0\rangle$ and $|1\rangle$ is given by the likelihood function
	\begin{equation}\label{eq:FBS}
		P(m|\varepsilon,\Updelta f,\tau)=\frac{1}{2}+\frac{m}{2}\{\alpha+\beta e^{-\tau/T}\cos{[2\pi(\Updelta f-\varepsilon)\tau]}\},
	\end{equation}
	where the parameters $\alpha = -0.02$ and $\beta = 0.6$ are coefficients that together capture state preparation and readout errors, while $T$ describes a coherence time that limits the duration of the evolution time $\tau$ (see the Supplemental Material~\cite{supplementary} for further details on how they impact the algorithm performance). We set $T=\SI{10}{\micro\second}$ based on the Hahn echo coherence time $T_{2\text{E}}\approx\SI{12}{\micro\second}$. We choose a value slightly lower than the estimated $T_{\text{2E}}$ to be on the safe side in case the intrinsic decoherence rate fluctuates on longer time scales. For reference, $T_2\approx \SI{42}{\micro\second}$ at $\Phi_{\text{ext}}=0$. The qubit fluctuation $\varepsilon$ is what we want the controller to estimate. 
	
	We emphasize that in this experiment the controller \emph{adaptively} computes in real time both the Ramsey evolution time $\tau$ and the drive frequency $f_\text{d}$ (and thus $\Updelta f$) for each probing cycle. This differs from previously employed approaches where (i) the evolution time $\tau$ is fixed~\cite{Vepsaelaeinen2022} and (ii) only either the evolution time $\tau$ \emph{or} the effective rotating frame frequency is chosen adaptively~\cite{Bonato2015, joas2021, Arshad2024, Berritta2024b}.
    
   Adaptive control of the phase theoretically allows for one bit of information to be gained per experiment in an error-free qubit. The estimation begins with the longest sensing time, which corresponds to the smallest frequency sensitivity~\cite{childs_quantum_2000}, and proceeds with shorter sensing times that yield larger frequency resolutions. To improve robustness against experimental errors, the strategy can be modified by repeating measurements with reduced information gain per measurement~\cite{said_nanoscale_magnetometry_2011, cappellaro_spin-bath_narrowing_2012}, which has been experimentally demonstrated using nitrogen-vacancy centers~\cite{nusran_high-dynamic-range_2012, waldherr_high-dynamic-range_2012, Bonato2015}. However, this phase estimation approach presents some limitations: (i) an extensive parameter search is required to determine the optimal number of repetitions for each sensing time~\cite{Bonato2015}, (ii) a Bayesian approach would yield a multi-modal probability distribution that requires many more than two parameters to be stored in memory, and (iii) the estimate becomes only useful after the very last cycle.

    We also highlight that using a fixed evolution time $\tau$ in non-adaptive probing cycles~\cite{Vepsaelaeinen2022} limits the frequency range to $\left(-\frac{1}{2\tau},+ \frac{1}{2\tau}\right]$, which results in a trade-off between range and frequency sensitivity. This trade-off does not apply to the binary search algorithm introduced here.
	
	In the quasistatic approximation, we want to find the optimal sequence of $\tau$'s and $f_\text{d}$'s to estimate $\varepsilon$ in as few measurements as possible. 
	Using Bayes' rule we write
	\begin{equation} \label{eq:posterior_FBS}
		{\cal P}_{n+1}(\varepsilon) \propto {\cal P}_n(\varepsilon)P(m_{n+1}|\varepsilon,\Updelta f_{n+1},\tau_{n+1}),
	\end{equation}
	where ${\cal P}_n(\varepsilon)$ denotes the probability distribution for $\varepsilon$ after the $n^{\text{th}}$ probing cycle, which thus depends on all previously used $\Updelta f_i$ and $\tau_i$ and all measurement outcomes $m_i$ (i.e., $i\leq n$).
	
	In the following we assume ${\cal P}_n(\varepsilon)$ to be a Gaussian, which well approximates the actual distribution (see the Supplemental Material~\cite{supplementary} for more details). The advantage of using the Gaussian approximation is that it allows the distribution to be conveniently described using just two parameters: the mean $\mu$ and the standard deviation $\sigma$.
	
	For each individual probing cycle, to minimize the expected posterior variance, in a greedy approach the optimal experiment $(\tau,\Updelta f)$ is the one that divides the prior distribution into equal (or approximately equal) left and right portions [see Fig.~\ref{fig:fig1_FBS}(b)] so that a half period of Eq.~\eqref{eq:FBS} is comparable to the width of the prior distribution, performing something similar to a binary search partitioning~\footnote{We clarify that this is not a true binary search in the sense of gaining exactly one bit of information per measurement. 
		The search does, however, follow a binary search tree where the two options at each node provide the most information within the approximations we use.
		A true \emph{quantum} binary search could be implemented for a decoherence-free qubit with ideal initialization and readout~\cite{childs_quantum_2000}. The focus here is on estimating noise in a physical qubit.}.
	The approach of using Gaussians and partitioning the posterior distributions in a similar fashion was introduced by Ref.~\cite{Ferrie2013}. 
	However, in that case, the lack of phase control in the likelihood prevents the algorithm from determining the sign of $\varepsilon$, making it ineffective for priors where $\mu/\sigma \rightarrow 0$~\footnote{In the limit $\mu/\sigma \rightarrow 0$, for $m = +1$, the likelihood function with zero phase (a squared cosine) has a single maximum at $\mu = 0$ within the 95\% credible interval of the prior distribution, and the corresponding posterior distribution can be approximated by a Gaussian. In contrast, for $m = -1$ in the same limit, the zero-phase likelihood function (a squared sine) exhibits two global maxima within the 95\% credible interval and a minimum at the maximum prior value, resulting in a bimodal posterior distribution that cannot be adequately approximated by a single Gaussian.}. While fitting to a bimodal Gaussian has been suggested as a solution~\cite{Benestad2024}, this approach still does not resolve the ambiguity of the sign. In this work, we address both aspects by dynamically updating $f_\text{d}$ (and thus $\Updelta f$) in real time on the controller. This allows the controller to consistently use the optimal likelihood function and perform the most efficient estimation using only a Gaussian distribution.
	
	Two steps are needed to implement the binary search in the controller. The first step is to determine optimal experiment parameters $\tau$ and $\Updelta f$ based on the prior distribution. Then, one must approximate the resulting posterior ${\cal P}_{n+1}(\varepsilon)$ to a Gaussian. This can be done using equations implemented directly on the controller in real time. 
	
	The prior distribution, dashed black line in Fig.~\ref{fig:fig1_FBS}(b), in the Gaussian approximation is given by
	\begin{equation}\label{eq:gauss_prior}
		{\cal P}_n(\varepsilon)=\frac{1}{\sqrt{2\pi \sigma_n^{2}}}\exp{-\left( \frac{\varepsilon-\mu_{n}}{\sqrt{2}\sigma_{n}}\right)^2 },
	\end{equation}
	where $\mu_n$ is our guess for what $\varepsilon$ is with uncertainty $\sigma_{n}$.
	In order to separate the prior at the center by the likelihood functions, the detuned frequency must satisfy
	\begin{equation}\label{eq:FBS_detuning}
		\Updelta f_{n+1}=\frac{1 + 2 l}{4\tau_{n+1}}+\mu_{n},
	\end{equation}
	where $l$ can be any integer number; we set it to be zero. While this is formally the optimal solution for $\alpha=0$, it is a reasonable approximation for the optimum at other realistic values of $\alpha$.
	Given this constraint, the optimal evolution time $\tau$ that minimizes the posterior variance is given by:
	\begin{equation}
		\tau_{n+1}=\frac{\sqrt{16\pi^2\sigma_n^2+1/T^2}-1/T}{8\pi^2\sigma_n^2}.
	\end{equation}
	After measuring $m_{n+1}=\pm 1$, the posterior distribution is obtained by using Eq.~(\ref{eq:posterior_FBS}), after which
	it is fit to a Gaussian using the method of moments, i.e., by calculating its mean and variance~\cite{Ferrie2013}
	\begin{subequations}
		\begin{align}
			\mu_{n+1} &=\mu_n-\frac{2\pi m_{n+1}\beta\sigma_n^2\tau_{n+1} e^{-\tau_{n+1}/T-2\pi^2\sigma_n^2\tau_{n+1}^2}}{1+m_{n+1}\alpha},\\
			\sigma_{n+1}^2& =\sigma_n^2-\frac{4\pi^2\beta^2\sigma_n^4\tau_{n+1}^2e^{-2\tau_{n+1}/T-4\pi^2\sigma_n^2\tau_{n+1}^2}}{(1+m_{n+1}\alpha)^2},\label{eq:FBS_update}
		\end{align}
	\end{subequations}
	and using these values to construct the new prior to be a Gaussian distribution with mean and variance $\mu_{n+1}$ and $\sigma^2_{n+1}$ [see Fig.~\ref{fig:fig1_FBS}(b)].
	This scheme is repeated $N$ times to obtain a sufficiently narrow distribution with exponential scaling with the number of measurements given by Eq.~(\ref{eq:FBS_update}) \footnote{Note that we do not claim exponential scaling with sensing time, which is fundamentally bounded by the Heisenberg limit $\sigma^2(\varepsilon) \sim 1/\tau^2$. Instead, our protocol achieves exponential scaling with the number of single-shot measurements $N$.}. 
	
	In Fig.~\ref{fig:fig1_FBS}(c) we illustrate the resulting evolution of ${\cal P}_n(\varepsilon)$ as a function of the measurement number $n$ for one representative estimation sequence, with an initial prior with $\sigma_0= \SI{1}{\mega\hertz}$ and $\mu_0 = 0$, and $N = 15$. The black dots show the expectation value $\langle \varepsilon \rangle$, and the shaded area indicates its 68\% credible interval, narrowing here down to $\SI{67}{\kilo\hertz}$ after 15 measurements in less than $\SI{100}{\micro\second}$. With similar parameters, this efficient frequency estimation could replace currently used periodic calibration routines by narrowing the qubit frequency within a user-defined error budget. As the FBS relies only on single-qubit operations, it can be extended to multi-qubit systems without introducing additional overhead.
	
	In the following, we optimize the number of single-shot measurements $N$ we use based on the specific experiment we are performing. We test the performance of the FBS by performing Ramsey repetitions to verify that it finds the correct qubit frequency, and we use randomized benchmarking and gate set tomography to assess the impact of the FBS on the qubit fidelity and drift. We emphasize again that the goal of these experiments is not to increase qubit coherence time or fidelity as much as possible, but simply to confirm that the information gained from the FBS is most likely correct.

	\section{Results}
	\subsection{Qubit coherence time $T_2^*$}
	
	We thus validate the FBS by rapid estimation of the shift of the qubit frequency $\varepsilon$ and demonstrating an extended coherence of the flux-tunable transmon qubit. 
	
	The fluctuating parameter $\varepsilon$ is estimated from the probing sequence shown in the top part of Fig.~\ref{fig:fig2_FBS}(a): For each probe cycle, the Bloch vector is positioned on the equator using an X$_{\pi/2}$ pulse. After precessing for a time $\tau_i$, a virtual Z$(\Updelta f_i)$ gate~\cite{McKay2017} is performed, introducing a phase offset $\phi = 2\pi\Updelta f_i \tau_i$. Here, virtual means that the phase offset is added to the subsequent X$_{\pi/2}$ pulse, which projects the Bloch vector back onto the $\hat{z}$-axis. 
    The virtual Z gate controls the phase $\Delta f \tau$ of the likelihood function [Eq.~(\ref{eq:FBS})] based on Eq.~(\ref{eq:FBS_detuning}).
	This is followed by a measurement, after which the qubit state $s_i$, ground ($s_i=0$) or excited ($s_i=1$), is assigned by thresholding the demodulated dispersive readout signal on the controller. 
	
	The Bayesian probability distribution of $\varepsilon$ is updated after comparing the measurement outcome $s_i$ to the previous one $s_{i-1}$, as we do not initialize the qubit to the ground state at the beginning of each cycle to reduce the probing cycle period (thus yielding higher feedback bandwidth). It follows that in Eq.~(\ref{eq:FBS}) $m_i =2|s_i - s_{i-1}| - 1$. For instance, if in the previous measurement the qubit was in the ground state $s_{i-1} = 0$ and now it is in the excited state ($s_{i} = 1$) then $m_i = +1$. The duration of the probe cycle is short compared to the measured $T_1\approx\SI{80}{\micro\second}$ at quarter flux: the average Ramsey evolution time is $\approx \SI{4.36}{\micro\second}$, the time used to read out the qubit is $\SI{1,44}{\micro\second}$, and the time used to subsequently cool down the resonator again (to deplete it from any residual photons after the readout pulse) is $\approx \SI{2}{\micro\second}$.
	
	The controller is programmed to start from a prior with $\mu_0 = 0$ and $\sigma_0 = \SI{30}{\kilo\hertz}$ based on previously measured qubit frequency fluctuations. After each measurement, the controller updates the probability distribution, resulting in a narrowing of the width of the distribution [cf.~Fig.~\ref{fig:fig1_FBS}(c)]. In the middle panel of Fig.~\ref{fig:fig2_FBS}(a) we plot the experimentally found final posterior probability distributions ${\cal P}_N(\varepsilon)$ from all estimations done in the full experiment, where we consistently use $N=8$ measurements per estimation sequence. 
	For plotting clarity we down-sample to $\approx \SI{14}{\milli\second}$.
	
	To confirm that these narrowed distribution functions indeed represent more accurate and correct knowledge about $\varepsilon$, we interleave the estimations with a series of Ramsey repetitions, half of them making use of the outcome of the estimations and half of them not, as shown schematically in the upper part of Fig.~\ref{fig:fig2_FBS}.
	After each estimation sequence of $\varepsilon$, the controller updates the qubit-frequency parameter $f_\text{q}$ (in software) to compensate for the estimated shift $\langle \varepsilon \rangle$, so that the total expected detuning becomes $\Updelta f_j - \langle \varepsilon\rangle = \SI{1}{\mega\hertz}$.
	Then it performs one Ramsey cycle with evolution time $\tau_j$, after which it again estimates $\varepsilon$. The intentional detuning of $\SI{1}{\mega\hertz}$ applied via the virtual Z gate improves the visibility of the Ramsey fringes, making them easier to fit.
	This cycle is repeated $M=50$ times, while $\tau_j$ is increased linearly from 0 to $\SI{7}{\micro\second}$.
	Each row in the middle panel of Fig.~\ref{fig:fig2_FBS}(b) shows the result of one such set of 50 Ramsey cycles, where we plot all single-shot measurement outcomes $p_j$ as white/black pixels.
	After each set of $M=50$ Ramsey cycles with feedback, the controller resets $f_\text{q}$ (in software) to the offline-calibrated value, tuning to $\Updelta f_k = \SI{1}{\mega\hertz}$.
	It then performs the same series of 50 Ramsey cycles as before, linearly stepping the evolution time $\tau_k$ from 0 to $\SI{7}{\micro\second}$.
	The rows in the middle panel of Fig.~\ref{fig:fig2_FBS}(c) show the single-shot measurement outcomes of this part of the protocol.
	After completing the Ramsey cycles without feedback, the whole protocol is repeated, as indicated in the top of Fig.~\ref{fig:fig2_FBS}, starting with a new estimation of $\varepsilon$. In the new estimation sequence, the controller starts from a prior with $\mu_0$ equal to the estimated $\varepsilon$ of the previous sequence.
	
	To highlight the effect of the feedback, we plot the averages of all Ramsey repetitions in the lower panels of Fig.~\ref{fig:fig2_FBS}(b,c).
	In both cases, we fit the decaying signal using Gaussian envelopes (solid line), yielding $T_2^*= \SI{3.73 (0.11)}{\micro\second}$ without feedback and $T_2^*= \SI{5.57 (0.09)}{\micro\second}$ with feedback, corresponding to a $\approx 49$\% improvement when including feedback.
	This increased coherence of the qubit demonstrates that the narrowed distribution function obtained from the estimation procedure indeed reflects more accurate knowledge about the fluctuating parameter $\varepsilon$. Notably, our approach requires fewer single-shot measurement outcomes ($N = 8$) compared to $N = 20$ of Ref.~\cite{Vepsaelaeinen2022}. The $T_2^*$ coherence time achieved with feedback is ultimately limited by the Hahn echo value due to the lower bandwidth of the feedback loop (approximately $\SI{16}{\kilo\hertz}$ for $N = 8$), compared to the much higher bandwidth of the dynamically decoupled Ramsey ($1/\tau \gtrapprox \SI{100}{\kilo\hertz}$). This explains why the Hahn echo is less sensitive to quasistatic fluctuations than the feedback-stabilized Ramsey experiment.

    We perform another experiment for 6 hours, from which we extract the noise power spectral density and show that the FBS keeps track of the qubit frequency also over longer time scales (see the Supplemental Material~\cite{supplementary}).
	Overall, the results presented in this section show how the FBS can efficiently find and stabilize the qubit frequency.
	
	\subsection{Single-qubit gate fidelity}
	\begin{figure}
		\centering
		\includegraphics{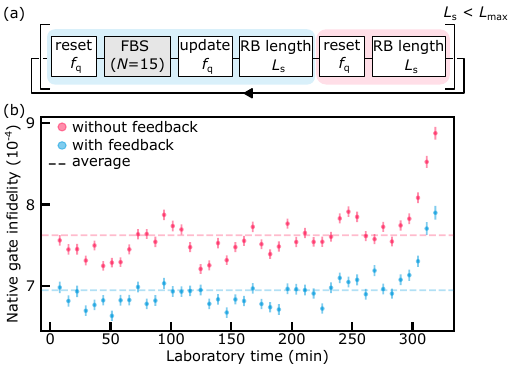}
		\caption{\textbf{Frequency binary search validation by randomized benchmarking}.
			\textbf{(a)} A randomized benchmarking repetition of length $L_{\text{s}}$ combined with the FBS feedback (with $N=15$ probing cycles). 
			\textbf{(b)} Randomized benchmarking is repeated 44 times with the feedback (blue dots) and without (red dots). The native gate errors are extracted from the fit to the data, which consists of 10,000 realizations of 30 random Clifford sequences with and without the feedback. The error bars show the 68\% confidence interval for the fitted gate error. Dashed lines are the averaged gate infidelities.
		}
		\label{fig:fig3_FBS}
	\end{figure}
	
	In this section we further validate the FBS calibration by showing improvement of the single-qubit gate fidelity by randomized benchmarking (RB)~\cite{Knill2008}. 
	
	The pulse sequence is shown in Fig.~\ref{fig:fig3_FBS}(a): the controller resets the qubit-frequency parameter $f_\text{q}$ (in software), and it performs the FBS with $N=15$ probing sequences, starting with $\mu_0=0$ and $\sigma_0 = \SI{200}{\kilo\hertz}$ \footnote{In the RB experiment, $\mu_0$ is reset to zero before each estimation sequence, which explains the increased $\sigma_0$ and $N$ compared to the Ramsey experiment of the previous section, where we do not reset $\mu_0$ before each estimation sequence. We believe that further improvements in the single-qubit gate fidelity could be achieved by using the same settings as in the Ramsey experiment. Still, this work focuses on demonstrating the enhancement provided by FBS, rather than achieving the highest possible fidelity for this particular setup.}. At the end of the estimation, $f_\text{q}$ is updated (in software) and an RB sequence of depth $L_{\text{s}}$ is performed. An interleaved RB measurement without feedback follows by resetting the qubit frequency to the offline-calibrated value. The maximum circuit depth ($L_{\text{max}}$) is 2,300, and the repetition is averaged 10,000 times. We implement DRAG (derivative reduction by adiabatic gate) for our $\SI{20}{\nano\second}$-long pulses to suppress leakage errors~\cite{Krantz2019, Blais2021} in the RB experiment. 
	As with the previous experiment, the qubit is not initialized in the ground state at the beginning of each cycle; the controller keeps track of whether the state is different or not compared to the previous measurement. Every 1,000 averages, the threshold that classifies the demodulated dispersive readout signal is updated online in the controller by taking the average of 10,000 single-shot measurements after performing an X$_{\pi/2}$ pulse to the qubit.
	
	The controller performs the RB experiment for 6 hours, yielding a native gate infidelity of $(7.6\pm0.3)\times 10^{-4}$ without feedback and $(6.9\pm0.2)\times 10^{-4}$ with feedback [dashed lines in Fig.~\ref{fig:fig3_FBS}(b)]. The single-qubit gate infidelities are higher than the decoherence limit~\cite{Malley2015} approximated by $t_{\text{gate}}(1/T_1 + 1/T_{\phi})/3\approx6\times10^{-4} $, given $t_{\text{gate}} \approx \SI{20}{\nano\second}$, $T_1\approx\SI{80}{\micro\second}$ and the exponential part of the pure dephasing time $T_{\phi}\approx \SI{13}{\micro\second}$. The feedback protocol always performs better than without feedback, and with less spread around the mean value as a result of the stabilization. Some of the drifts of the infidelity remain correlated, which we tentatively attribute to other factors (e.g., changes in $T_1$). The improved fidelities of single-qubit gates by feedback by interleaved RB are presented in the Supplemental Material~\cite{supplementary}. 
	We attribute the smaller relative improvement in the single-qubit gate qubit fidelity, compared to coherence in the previous section, to the larger final value of \( \sigma_{N=15} \approx \SI{90}{\kilo\hertz} \), as opposed to \( \sigma_{N=8} \approx \SI{24}{\kilo\hertz} \), which resulted from different initial prior distributions.
	
	\subsection{Reduction of non-Markovian noise}
	\begin{figure*}
		\centering
		\includegraphics{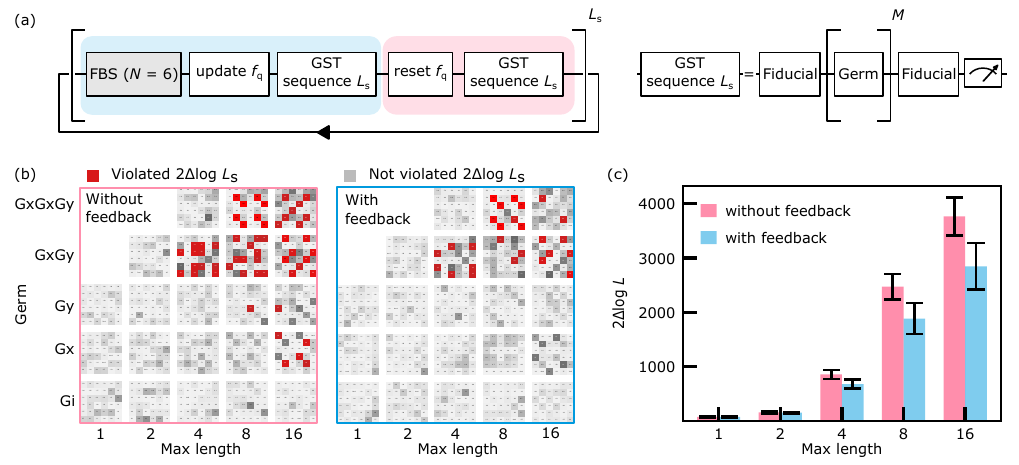}
		\caption{\textbf{Gate set tomography and model violation}.
			\textbf{(a)} A gate set tomography (GST) sequence combined with the FBS feedback (with $N=6$ probing cycles).
			\textbf{(b)} Model violation plot without (left) and with (right) FBS feedback. The red marks reveal detection of model violation at a confidence level of more than 95\% and the gray boxes indicate statistical fluctuations.
			\textbf{(c)} The FBS feedback decreases the total amount of the log-likelihood ratios $2\Updelta \log L$ at maximum lengths of 4, 8, and 16, extracted from 30 GST runs. Error bars show the 68\% credible interval.
		}
		\label{fig:fig4_FBS}
	\end{figure*}
	
	So far, we have demonstrated an efficient qubit calibration protocol by instructing a controller to generate adaptive probing sequences in real time. The calibration has been validated by improved coherence and fidelity. Next, we investigate if our estimation protocol also reduces the non-Markovian noise in the system, this time estimated by gate set tomography (GST)~\cite{Nielsen2021}. While GST has been employed to validate the mitigation of drifts in semiconductor spin qubits~\cite{Park2025} by real-time estimation, we present its application for validating drift mitigation in a superconducting qubit by real-time feedback.
    
    GST is a calibration-free method for benchmarking and characterizing operations in a quantum processor and we implement it by using the pyGSTi~\cite{Nielsen2020} software package. GST relies on running circuits designed to amplify certain types of errors. Each circuit, as shown in the right panel of Fig.~\ref{fig:fig4_FBS}(a), consists of the gate to be characterized, $\{G_\text{I}, G_\text{x}, G_\text{y}, G_\text{x} G_\text{y},G_\text{x}G_\text{x}G_\text{y} \}$ (the germs), sandwiched between two fiducial circuits from the set $\{\text{I}, \text{X}_{\pi/2}, \text{Y}_{\pi/2}, \text{X}_{\pi/2}\text{X}_{\pi/2}, \text{X}_{\pi/2}\text{X}_{\pi/2}\text{X}_{\pi/2}, \text{Y}_{\pi/2}\text{Y}_{\pi/2}\text{Y}_{\pi/2} \} $. These are applied to the qubit after initialization, to generate the state $\rho_i$, and before the measurement in the computational basis, to measure the operator $E_j$. The variable $L_\text{s}$ is the maximum depth used to construct the base circuit from the germ~\cite{Nielsen2021}. By running the circuit multiple times, we extract the measurement probabilities \( P_{ji} = \Tr{E_j G[\rho_i]}\), which are then fitted to the error model, providing a maximum likelihood estimator of the transfer matrix for all of the gates. The quality of the fit can be used to detect violations of the Markovian model of the noise~\cite{Nielsen2021, hashim2024practical}.

	We incorporate the feedback loop inside the GST protocol using the scheme shown in the left panel of Fig.~\ref{fig:fig4_FBS}(a), implemented on the controller. We find suitable parameters for testing the FBS to be \( N=6 \) per estimation sequence with feedback, with initial \( \mu_0=0 \) and \( \sigma_0= \SI{30}{\kilo\hertz} \). Each repetition is executed 100 times for each of the 616 sequences constructed for GST. In the new estimation sequence, the controller starts from a prior with $\mu_0$ equal to the estimated $\varepsilon$ of the previous repetition.
	
	The measured discrepancy between the Markovian model and the data ($2\Updelta \log L_\text{s}$ as defined in Ref.~\cite{Nielsen2021}) is shown in Fig.~\ref{fig:fig4_FBS}(b) using the color of the squares: The grayscale is used to reflect the total discrepancy and a red color signals a model violation, indicating non-Markovian noise at 95\% confidence interval. The rows and columns of the inner matrices represent the fiducial operations used for state preparation and measurement respectively. For the data with feedback, represented in the right panel of Fig.~\ref{fig:fig4_FBS}(b), a clear decrease in the number of red squares in comparison to the feedback-free result of the left panel is visible.
	
	To further quantify the reduction of non-Markovian noise, the controller repeats the GST experiment 30 times, spanning a laboratory time of 40 minutes. In Fig.~\ref{fig:fig4_FBS}(c), we plot, as a function of the maximum length, the total violation of the model. While there is no clear improvement for maximum lengths of 1 and 2, it is evident for longer sequences. As shown in the Supplementary Material~\cite{supplementary}, feedback allows to significantly decrease the amplitude of non-Markovian noise, identified as frequency fluctuations below the inverse of feedback timescale. It complements the previous analysis of gate set tomography data, which is consistent with the expectation that longer circuits are more sensitive to non-Markovian noise~\cite{Nielsen2021}. We attribute the residual discrepancy between the model and the data, especially where the $G_{\text{x}}$ germ is used, to larger $|\varepsilon|$ fluctuations not captured by the chosen initial prior distribution $\mathcal{P}_0(\varepsilon)$. Further improvements would require a more quantitative analysis of the trade-off between the initially chosen uncertainty, the estimation bandwidth, and the white noise floor of the estimation method (see the Supplemental Material~\cite{supplementary}).
	
	Regarding gate performance, the GST protocol yields an X\(_{\pi/2} \)-pulse infidelity of \( (2.6\pm0.3)\times 10^{-2} \) and for the Y\(_{\pi/2} \)-pulse \( (1.6\pm0.3)\times 10^{-2} \), both with and without feedback. These values are two orders of magnitude larger than those reported by RB, which further suggests the dominant role of coherent noise, to which RB is less sensitive~\cite{Sanders2015}. In summary, the decrease in Markovian model violation for longer GST sequences indicates that the FBS feedback protocol mitigates a significant portion of non-Markovian noise.
	\section{Outlook}
	
	Our work presents the experimental demonstration of a greedy adaptive Bayesian estimation protocol for the frequency of a resonantly driven qubit, improving its performance using only $N= 6$ single-shot measurements leveraging the exponential scaling of the algorithm with the number of measurements. This small number is to be compared with the tens~\cite{Vepsaelaeinen2022} or hundreds of single-shot measurements of previous works~\cite{Gebhart2023, Nakajima2020, Park2025} in resonantly driven qubits.
	Our approach reduces qubit-frequency drift by real-time Bayesian estimation and feedback, implemented via a low-latency FPGA-based qubit control system.
	
	The binary search estimation algorithm allows control pulses to compensate for qubit frequency fluctuations caused by flux noise, improving coherence and fidelity without sacrificing frequency sensitivity or range. We validate the protocol using gate set tomography, which further corroborates our claim that our adaptive feedback loop reduces the effects of non-Markovian noise. This may facilitate quantum error correction methods, which generally assume Markovian noise. 
		
	Our protocol assumes that a Gaussian distribution can approximate the frequency fluctuations and that these remain constant over the time scale of a few measurements. In this work, the controller updates the drive frequency in real time, but the scheme can be adapted to adjust the qubit frequency by modifying the flux bias instead. The estimation bandwidth is limited by the relatively slow measurement time of a few microseconds. From this perspective, our work represents a worst-case scenario, underlying the efficacy of our experimental technique.

    Traditional real-time approaches for continuous parameter estimations have relied on simple linear or trigonometric functions or precomputed lookup tables~\cite{Gebhart2023, Nakajima2020, Arshad2024, Berritta2024a, Berritta2024b, Park2025}. In contrast, our FPGA implementation performs real-time Bayesian inference and processes widely used Gaussian distributions, resulting in a numerical complexity higher than previous works. At the same time, the parametrization of Gaussian distributions allows the FPGA to perform fewer and faster computations compared to traditional approaches using histograms and particle filtering.
	
	We anticipate further improvements possibly by the implementation of a physics-informed prior~\cite{Berritta2024b} that accounts for $1/f$ noise. Also, the online calibration of single-qubit gates is a possible direction with available hardware, for single-qubit corrections in two-qubit gates by actual (not virtual) Z gates. 
    
    Adaptive Bayesian techniques could be implemented for frequency tracking in the presence of two-level fluctuators~\cite{Liu2024, Ye2024}. Specific to our flux-tunable transmon qubit, while the feedback bandwidth is limited by readout and resonator cooldown time, higher bandwidth can be achieved by adding a Purcell filter~\cite{Krantz2019, Blais2021} which protects the qubit from relaxing into its environment. Our protocol may favor using symmetric junctions to increase the frequency range of the qubit, without worrying about increased sensitivity to flux noise as shown in Ref.~\cite{Vepsaelaeinen2022}.
	
	Although qubit calibration by Bayesian inference can be relatively easily integrated with existing classical control hardware, it requires prior knowledge of the system’s parametrization. We expect that quantum model learning agents~\cite{Gebhart2023}, which are more challenging to implement efficiently in real time, will gain wider adoption. We anticipate further research that merges theoretical and hardware advances, ultimately eliminating the need for a case-to-case modeling of experimental parameters.
	
	Beyond superconducting qubits, our scheme offers new insights into real-time calibration of any qubits manipulated by resonant pulses. The protocol is not restricted to flux noise but is compatible with any source of low-frequency noise. We envision future applications in mitigating quasistatic electrical and nuclear noise in other solid-state qubit platforms.
    
    This work advances quantum control by implementing an adaptive Bayesian technique to calibrate the qubits frequencies in real time. Our algorithm is a locally optimal solution by minimizing the expected estimator variance under a Gaussian distribution approximation, making it appealing for real-time calibration in large QPUs.

	\section{Author contributions}
	FB, JB, and JAK conceptualized the experiment. FB led the measurements and data analysis, and wrote the manuscript with input from all authors. FB, LP, MM, RA, AC, JAG, WDO, and FK performed the experiment with theoretical contributions from JB, JAK, and JD. YS designed the device, which was fabricated by DKK and BMN under supervision of KS, MES, and JLY. JAG, JD, WDO, and FK supervised the project.
	\section{Acknowledgments}
	We gratefully acknowledge Patrick Harrington and Max Hays for fruitful discussions.
	This work received funding from the European Union’s Horizon 2020 research and innovation programme under grant agreements 101017733 (QuantERA II), 951852 (QLSI), EUREKA Eurostars 3 (ECHIDNA), European Research Council (ERC grant 856526), the Novo Nordisk Foundation under Challenge Programme NNF20OC0060019 (SolidQ), the Inge Lehmann Programme of the Independent Research Fund Denmark, the US Army Research Office (ARO) under Award No. W911NF-24-2-0043, the QuSpin Mobility Grant, the INTFELLES-Project No. 333990, which is funded by the Research Council of Norway (RCN), the Dutch National Growth Fund (NGF) as part of the Quantum Delta NL programme, and from the Danish Agency for Higher Education and Science (DAHES, grant 2076-00014B). It was also funded in part by the US Army Research Office (ARO) Multidisciplinary University Research Initiative (MURI) W911NF-18-1-0218, in part by the US Army Research Office (ARO) under Award No. W911NF-23-1-0045, and in part under Air Force Contract No. FA8702-15-D-0001. Any opinions, findings, conclusions or recommendations expressed in this material are those of the author(s) and do not necessarily reflect the views of the US Air Force or the US Government.
	
	\appendix*

	\nocite{*}
	\bibliography{my_bibliography}

\begin{thebibliography}{57}%
\makeatletter
\providecommand \@ifxundefined [1]{%
 \@ifx{#1\undefined}
}%
\providecommand \@ifnum [1]{%
 \ifnum #1\expandafter \@firstoftwo
 \else \expandafter \@secondoftwo
 \fi
}%
\providecommand \@ifx [1]{%
 \ifx #1\expandafter \@firstoftwo
 \else \expandafter \@secondoftwo
 \fi
}%
\providecommand \natexlab [1]{#1}%
\providecommand \enquote  [1]{``#1''}%
\providecommand \bibnamefont  [1]{#1}%
\providecommand \bibfnamefont [1]{#1}%
\providecommand \citenamefont [1]{#1}%
\providecommand \href@noop [0]{\@secondoftwo}%
\providecommand \href [0]{\begingroup \@sanitize@url \@href}%
\providecommand \@href[1]{\@@startlink{#1}\@@href}%
\providecommand \@@href[1]{\endgroup#1\@@endlink}%
\providecommand \@sanitize@url [0]{\catcode `\\12\catcode `\$12\catcode
  `\&12\catcode `\#12\catcode `\^12\catcode `\_12\catcode `\%12\relax}%
\providecommand \@@startlink[1]{}%
\providecommand \@@endlink[0]{}%
\providecommand \url  [0]{\begingroup\@sanitize@url \@url }%
\providecommand \@url [1]{\endgroup\@href {#1}{\urlprefix }}%
\providecommand \urlprefix  [0]{URL }%
\providecommand \Eprint [0]{\href }%
\providecommand \doibase [0]{https://doi.org/}%
\providecommand \selectlanguage [0]{\@gobble}%
\providecommand \bibinfo  [0]{\@secondoftwo}%
\providecommand \bibfield  [0]{\@secondoftwo}%
\providecommand \translation [1]{[#1]}%
\providecommand \BibitemOpen [0]{}%
\providecommand \bibitemStop [0]{}%
\providecommand \bibitemNoStop [0]{.\EOS\space}%
\providecommand \EOS [0]{\spacefactor3000\relax}%
\providecommand \BibitemShut  [1]{\csname bibitem#1\endcsname}%
\let\auto@bib@innerbib\@empty
\bibitem [{\citenamefont {Ichikawa}\ \emph {et~al.}(2024)\citenamefont
  {Ichikawa}, \citenamefont {Hakoshima}, \citenamefont {Inui}, \citenamefont
  {Ito}, \citenamefont {Matsuda}, \citenamefont {Mitarai}, \citenamefont
  {Miyamoto}, \citenamefont {Mizukami}, \citenamefont {Mizuta}, \citenamefont
  {Mori}, \citenamefont {Nakano}, \citenamefont {Nakayama}, \citenamefont
  {Okada}, \citenamefont {Sugimoto}, \citenamefont {Takahira}, \citenamefont
  {Takemori}, \citenamefont {Tsukano}, \citenamefont {Ueda}, \citenamefont
  {Watanabe}, \citenamefont {Yoshida},\ and\ \citenamefont
  {Fujii}}]{Ichikawa2024}%
  \BibitemOpen
  \bibfield  {author} {\bibinfo {author} {\bibfnamefont {T.}~\bibnamefont
  {Ichikawa}}, \bibinfo {author} {\bibfnamefont {H.}~\bibnamefont {Hakoshima}},
  \bibinfo {author} {\bibfnamefont {K.}~\bibnamefont {Inui}}, \bibinfo {author}
  {\bibfnamefont {K.}~\bibnamefont {Ito}}, \bibinfo {author} {\bibfnamefont
  {R.}~\bibnamefont {Matsuda}}, \bibinfo {author} {\bibfnamefont
  {K.}~\bibnamefont {Mitarai}}, \bibinfo {author} {\bibfnamefont
  {K.}~\bibnamefont {Miyamoto}}, \bibinfo {author} {\bibfnamefont
  {W.}~\bibnamefont {Mizukami}}, \bibinfo {author} {\bibfnamefont
  {K.}~\bibnamefont {Mizuta}}, \bibinfo {author} {\bibfnamefont
  {T.}~\bibnamefont {Mori}}, \bibinfo {author} {\bibfnamefont {Y.}~\bibnamefont
  {Nakano}}, \bibinfo {author} {\bibfnamefont {A.}~\bibnamefont {Nakayama}},
  \bibinfo {author} {\bibfnamefont {K.~N.}\ \bibnamefont {Okada}}, \bibinfo
  {author} {\bibfnamefont {T.}~\bibnamefont {Sugimoto}}, \bibinfo {author}
  {\bibfnamefont {S.}~\bibnamefont {Takahira}}, \bibinfo {author}
  {\bibfnamefont {N.}~\bibnamefont {Takemori}}, \bibinfo {author}
  {\bibfnamefont {S.}~\bibnamefont {Tsukano}}, \bibinfo {author} {\bibfnamefont
  {H.}~\bibnamefont {Ueda}}, \bibinfo {author} {\bibfnamefont {R.}~\bibnamefont
  {Watanabe}}, \bibinfo {author} {\bibfnamefont {Y.}~\bibnamefont {Yoshida}},\
  and\ \bibinfo {author} {\bibfnamefont {K.}~\bibnamefont {Fujii}},\ }\bibfield
   {title} {\bibinfo {title} {Current numbers of qubits and their uses},\
  }\href {https://doi.org/10.1038/s42254-024-00725-0} {\bibfield  {journal}
  {\bibinfo  {journal} {Nature Reviews Physics}\ }\textbf {\bibinfo {volume}
  {6}},\ \bibinfo {pages} {345} (\bibinfo {year} {2024})}\BibitemShut {NoStop}%
\bibitem [{\citenamefont {Campbell}(2024)}]{Campbell2024}%
  \BibitemOpen
  \bibfield  {author} {\bibinfo {author} {\bibfnamefont {E.}~\bibnamefont
  {Campbell}},\ }\bibfield  {title} {\bibinfo {title} {A series of fast-paced
  advances in quantum error correction},\ }\href
  {https://doi.org/https://doi.org/10.1038/s42254-024-00706-3} {\bibfield
  {journal} {\bibinfo  {journal} {Nature Reviews Physics}\ }\textbf {\bibinfo
  {volume} {6}},\ \bibinfo {pages} {160} (\bibinfo {year} {2024})}\BibitemShut
  {NoStop}%
\bibitem [{\citenamefont {Acharya}\ \emph {et~al.}(2024)\citenamefont
  {Acharya}, \citenamefont {Aghababaie-Beni}, \citenamefont {Aleiner},
  \citenamefont {Andersen}, \citenamefont {Ansmann}, \citenamefont {Arute},
  \citenamefont {Arya}, \citenamefont {Asfaw}, \citenamefont {Astrakhantsev},
  \citenamefont {Atalaya} \emph {et~al.}}]{Acharya2024}%
  \BibitemOpen
  \bibfield  {author} {\bibinfo {author} {\bibfnamefont {R.}~\bibnamefont
  {Acharya}}, \bibinfo {author} {\bibfnamefont {L.}~\bibnamefont
  {Aghababaie-Beni}}, \bibinfo {author} {\bibfnamefont {I.}~\bibnamefont
  {Aleiner}}, \bibinfo {author} {\bibfnamefont {T.~I.}\ \bibnamefont
  {Andersen}}, \bibinfo {author} {\bibfnamefont {M.}~\bibnamefont {Ansmann}},
  \bibinfo {author} {\bibfnamefont {F.}~\bibnamefont {Arute}}, \bibinfo
  {author} {\bibfnamefont {K.}~\bibnamefont {Arya}}, \bibinfo {author}
  {\bibfnamefont {A.}~\bibnamefont {Asfaw}}, \bibinfo {author} {\bibfnamefont
  {N.}~\bibnamefont {Astrakhantsev}}, \bibinfo {author} {\bibfnamefont
  {J.}~\bibnamefont {Atalaya}}, \emph {et~al.},\ }\bibfield  {title} {\bibinfo
  {title} {Quantum error correction below the surface code threshold}\ }\href
  {https://doi.org/10.48550/arXiv.2408.13687} {10.48550/arXiv.2408.13687}
  (\bibinfo {year} {2024})\BibitemShut {NoStop}%
\bibitem [{\citenamefont {Ball}\ \emph {et~al.}(2016)\citenamefont {Ball},
  \citenamefont {Oliver},\ and\ \citenamefont {Biercuk}}]{Ball2016}%
  \BibitemOpen
  \bibfield  {author} {\bibinfo {author} {\bibfnamefont {H.}~\bibnamefont
  {Ball}}, \bibinfo {author} {\bibfnamefont {W.~D.}\ \bibnamefont {Oliver}},\
  and\ \bibinfo {author} {\bibfnamefont {M.~J.}\ \bibnamefont {Biercuk}},\
  }\bibfield  {title} {\bibinfo {title} {The role of master clock stability in
  quantum information processing},\ }\href
  {https://doi.org/https://doi.org/10.1038/npjqi.2016.33} {\bibfield  {journal}
  {\bibinfo  {journal} {npj Quantum Information}\ }\textbf {\bibinfo {volume}
  {2}},\ \bibinfo {pages} {1} (\bibinfo {year} {2016})}\BibitemShut {NoStop}%
\bibitem [{\citenamefont {Mohseni}\ \emph {et~al.}(2024)\citenamefont
  {Mohseni}, \citenamefont {Scherer}, \citenamefont {Johnson}, \citenamefont
  {Wertheim}, \citenamefont {Otten}, \citenamefont {Aadit}, \citenamefont
  {Bresniker}, \citenamefont {Camsari}, \citenamefont {Chapman}, \citenamefont
  {Chatterjee}, \citenamefont {Dagnew}, \citenamefont {Esposito}, \citenamefont
  {Fahim}, \citenamefont {Fiorentino}, \citenamefont {Khalid}, \citenamefont
  {Kong}, \citenamefont {Kulchytskyy}, \citenamefont {Li}, \citenamefont
  {Lott}, \citenamefont {Markov}, \citenamefont {McDermott}, \citenamefont
  {Pedretti}, \citenamefont {Gajjar}, \citenamefont {Silva}, \citenamefont
  {Sorebo}, \citenamefont {Spentzouris}, \citenamefont {Steiner}, \citenamefont
  {Torosov}, \citenamefont {Venturelli}, \citenamefont {Visser}, \citenamefont
  {Webb}, \citenamefont {Zhan}, \citenamefont {Cohen}, \citenamefont {Ronagh},
  \citenamefont {Ho}, \citenamefont {Beausoleil},\ and\ \citenamefont
  {Martinis}}]{mohseni2024}%
  \BibitemOpen
  \bibfield  {author} {\bibinfo {author} {\bibfnamefont {M.}~\bibnamefont
  {Mohseni}}, \bibinfo {author} {\bibfnamefont {A.}~\bibnamefont {Scherer}},
  \bibinfo {author} {\bibfnamefont {K.~G.}\ \bibnamefont {Johnson}}, \bibinfo
  {author} {\bibfnamefont {O.}~\bibnamefont {Wertheim}}, \bibinfo {author}
  {\bibfnamefont {M.}~\bibnamefont {Otten}}, \bibinfo {author} {\bibfnamefont
  {N.~A.}\ \bibnamefont {Aadit}}, \bibinfo {author} {\bibfnamefont {K.~M.}\
  \bibnamefont {Bresniker}}, \bibinfo {author} {\bibfnamefont {K.~Y.}\
  \bibnamefont {Camsari}}, \bibinfo {author} {\bibfnamefont {B.}~\bibnamefont
  {Chapman}}, \bibinfo {author} {\bibfnamefont {S.}~\bibnamefont {Chatterjee}},
  \bibinfo {author} {\bibfnamefont {G.~A.}\ \bibnamefont {Dagnew}}, \bibinfo
  {author} {\bibfnamefont {A.}~\bibnamefont {Esposito}}, \bibinfo {author}
  {\bibfnamefont {F.}~\bibnamefont {Fahim}}, \bibinfo {author} {\bibfnamefont
  {M.}~\bibnamefont {Fiorentino}}, \bibinfo {author} {\bibfnamefont
  {A.}~\bibnamefont {Khalid}}, \bibinfo {author} {\bibfnamefont
  {X.}~\bibnamefont {Kong}}, \bibinfo {author} {\bibfnamefont {B.}~\bibnamefont
  {Kulchytskyy}}, \bibinfo {author} {\bibfnamefont {R.}~\bibnamefont {Li}},
  \bibinfo {author} {\bibfnamefont {P.~A.}\ \bibnamefont {Lott}}, \bibinfo
  {author} {\bibfnamefont {I.~L.}\ \bibnamefont {Markov}}, \bibinfo {author}
  {\bibfnamefont {R.~F.}\ \bibnamefont {McDermott}}, \bibinfo {author}
  {\bibfnamefont {G.}~\bibnamefont {Pedretti}}, \bibinfo {author}
  {\bibfnamefont {A.}~\bibnamefont {Gajjar}}, \bibinfo {author} {\bibfnamefont
  {A.}~\bibnamefont {Silva}}, \bibinfo {author} {\bibfnamefont
  {J.}~\bibnamefont {Sorebo}}, \bibinfo {author} {\bibfnamefont
  {P.}~\bibnamefont {Spentzouris}}, \bibinfo {author} {\bibfnamefont
  {Z.}~\bibnamefont {Steiner}}, \bibinfo {author} {\bibfnamefont
  {B.}~\bibnamefont {Torosov}}, \bibinfo {author} {\bibfnamefont
  {D.}~\bibnamefont {Venturelli}}, \bibinfo {author} {\bibfnamefont {R.~J.}\
  \bibnamefont {Visser}}, \bibinfo {author} {\bibfnamefont {Z.}~\bibnamefont
  {Webb}}, \bibinfo {author} {\bibfnamefont {X.}~\bibnamefont {Zhan}}, \bibinfo
  {author} {\bibfnamefont {Y.}~\bibnamefont {Cohen}}, \bibinfo {author}
  {\bibfnamefont {P.}~\bibnamefont {Ronagh}}, \bibinfo {author} {\bibfnamefont
  {A.}~\bibnamefont {Ho}}, \bibinfo {author} {\bibfnamefont {R.~G.}\
  \bibnamefont {Beausoleil}},\ and\ \bibinfo {author} {\bibfnamefont {J.~M.}\
  \bibnamefont {Martinis}},\ }\bibfield  {title} {\bibinfo {title} {How to
  build a quantum supercomputer: Scaling challenges and opportunities}\ }\href
  {https://doi.org/10.48550/arXiv.2411.10406} {10.48550/arXiv.2411.10406}
  (\bibinfo {year} {2024})\BibitemShut {NoStop}%
\bibitem [{\citenamefont {Gebhart}\ \emph {et~al.}(2023)\citenamefont
  {Gebhart}, \citenamefont {Santagati}, \citenamefont {Gentile}, \citenamefont
  {Gauger}, \citenamefont {Craig}, \citenamefont {Ares}, \citenamefont
  {Banchi}, \citenamefont {Marquardt}, \citenamefont {Pezz{\`e}},\ and\
  \citenamefont {Bonato}}]{Gebhart2023}%
  \BibitemOpen
  \bibfield  {author} {\bibinfo {author} {\bibfnamefont {V.}~\bibnamefont
  {Gebhart}}, \bibinfo {author} {\bibfnamefont {R.}~\bibnamefont {Santagati}},
  \bibinfo {author} {\bibfnamefont {A.~A.}\ \bibnamefont {Gentile}}, \bibinfo
  {author} {\bibfnamefont {E.~M.}\ \bibnamefont {Gauger}}, \bibinfo {author}
  {\bibfnamefont {D.}~\bibnamefont {Craig}}, \bibinfo {author} {\bibfnamefont
  {N.}~\bibnamefont {Ares}}, \bibinfo {author} {\bibfnamefont {L.}~\bibnamefont
  {Banchi}}, \bibinfo {author} {\bibfnamefont {F.}~\bibnamefont {Marquardt}},
  \bibinfo {author} {\bibfnamefont {L.}~\bibnamefont {Pezz{\`e}}},\ and\
  \bibinfo {author} {\bibfnamefont {C.}~\bibnamefont {Bonato}},\ }\bibfield
  {title} {\bibinfo {title} {Learning quantum systems},\ }\href
  {https://doi.org/https://doi.org/10.1038/s42254-022-00552-1} {\bibfield
  {journal} {\bibinfo  {journal} {Nature Reviews Physics}\ }\textbf {\bibinfo
  {volume} {5}},\ \bibinfo {pages} {141} (\bibinfo {year} {2023})}\BibitemShut
  {NoStop}%
\bibitem [{\citenamefont {Reuer}\ \emph {et~al.}(2023)\citenamefont {Reuer},
  \citenamefont {Landgraf}, \citenamefont {F{\"o}sel}, \citenamefont
  {O’Sullivan}, \citenamefont {Beltr{\'a}n}, \citenamefont {Akin},
  \citenamefont {Norris}, \citenamefont {Remm}, \citenamefont {Kerschbaum},
  \citenamefont {Besse} \emph {et~al.}}]{reuer2023realizing}%
  \BibitemOpen
  \bibfield  {author} {\bibinfo {author} {\bibfnamefont {K.}~\bibnamefont
  {Reuer}}, \bibinfo {author} {\bibfnamefont {J.}~\bibnamefont {Landgraf}},
  \bibinfo {author} {\bibfnamefont {T.}~\bibnamefont {F{\"o}sel}}, \bibinfo
  {author} {\bibfnamefont {J.}~\bibnamefont {O’Sullivan}}, \bibinfo {author}
  {\bibfnamefont {L.}~\bibnamefont {Beltr{\'a}n}}, \bibinfo {author}
  {\bibfnamefont {A.}~\bibnamefont {Akin}}, \bibinfo {author} {\bibfnamefont
  {G.~J.}\ \bibnamefont {Norris}}, \bibinfo {author} {\bibfnamefont
  {A.}~\bibnamefont {Remm}}, \bibinfo {author} {\bibfnamefont {M.}~\bibnamefont
  {Kerschbaum}}, \bibinfo {author} {\bibfnamefont {J.-C.}\ \bibnamefont
  {Besse}}, \emph {et~al.},\ }\bibfield  {title} {\bibinfo {title} {Realizing a
  deep reinforcement learning agent for real-time quantum feedback},\ }\href
  {https://doi.org/10.1038/s41467-023-42901-3} {\bibfield  {journal} {\bibinfo
  {journal} {Nature Communications}\ }\textbf {\bibinfo {volume} {14}},\
  \bibinfo {pages} {7138} (\bibinfo {year} {2023})}\BibitemShut {NoStop}%
\bibitem [{\citenamefont {Arshad}\ \emph {et~al.}(2024)\citenamefont {Arshad},
  \citenamefont {Bekker}, \citenamefont {Haylock}, \citenamefont {Skrzypczak},
  \citenamefont {White}, \citenamefont {Griffiths}, \citenamefont {Gore},
  \citenamefont {Morley}, \citenamefont {Salter}, \citenamefont {Smith},
  \citenamefont {Zohar}, \citenamefont {Finkler}, \citenamefont {Altmann},
  \citenamefont {Gauger},\ and\ \citenamefont {Bonato}}]{Arshad2024}%
  \BibitemOpen
  \bibfield  {author} {\bibinfo {author} {\bibfnamefont {M.~J.}\ \bibnamefont
  {Arshad}}, \bibinfo {author} {\bibfnamefont {C.}~\bibnamefont {Bekker}},
  \bibinfo {author} {\bibfnamefont {B.}~\bibnamefont {Haylock}}, \bibinfo
  {author} {\bibfnamefont {K.}~\bibnamefont {Skrzypczak}}, \bibinfo {author}
  {\bibfnamefont {D.}~\bibnamefont {White}}, \bibinfo {author} {\bibfnamefont
  {B.}~\bibnamefont {Griffiths}}, \bibinfo {author} {\bibfnamefont
  {J.}~\bibnamefont {Gore}}, \bibinfo {author} {\bibfnamefont {G.~W.}\
  \bibnamefont {Morley}}, \bibinfo {author} {\bibfnamefont {P.}~\bibnamefont
  {Salter}}, \bibinfo {author} {\bibfnamefont {J.}~\bibnamefont {Smith}},
  \bibinfo {author} {\bibfnamefont {I.}~\bibnamefont {Zohar}}, \bibinfo
  {author} {\bibfnamefont {A.}~\bibnamefont {Finkler}}, \bibinfo {author}
  {\bibfnamefont {Y.}~\bibnamefont {Altmann}}, \bibinfo {author} {\bibfnamefont
  {E.~M.}\ \bibnamefont {Gauger}},\ and\ \bibinfo {author} {\bibfnamefont
  {C.}~\bibnamefont {Bonato}},\ }\bibfield  {title} {\bibinfo {title}
  {Real-time adaptive estimation of decoherence timescales for a single
  qubit},\ }\href {https://doi.org/10.1103/physrevapplied.21.024026} {\bibfield
   {journal} {\bibinfo  {journal} {Physical Review Applied}\ }\textbf {\bibinfo
  {volume} {21}},\ \bibinfo {pages} {024026} (\bibinfo {year}
  {2024})}\BibitemShut {NoStop}%
\bibitem [{\citenamefont {Berritta}\ \emph
  {et~al.}(2024{\natexlab{a}})\citenamefont {Berritta}, \citenamefont
  {Rasmussen}, \citenamefont {Krzywda}, \citenamefont {van~der Heijden},
  \citenamefont {Fedele}, \citenamefont {Fallahi}, \citenamefont {Gardner},
  \citenamefont {Manfra}, \citenamefont {van Nieuwenburg}, \citenamefont
  {Danon}, \citenamefont {Chatterjee},\ and\ \citenamefont
  {Kuemmeth}}]{Berritta2024a}%
  \BibitemOpen
  \bibfield  {author} {\bibinfo {author} {\bibfnamefont {F.}~\bibnamefont
  {Berritta}}, \bibinfo {author} {\bibfnamefont {T.}~\bibnamefont {Rasmussen}},
  \bibinfo {author} {\bibfnamefont {J.~A.}\ \bibnamefont {Krzywda}}, \bibinfo
  {author} {\bibfnamefont {J.}~\bibnamefont {van~der Heijden}}, \bibinfo
  {author} {\bibfnamefont {F.}~\bibnamefont {Fedele}}, \bibinfo {author}
  {\bibfnamefont {S.}~\bibnamefont {Fallahi}}, \bibinfo {author} {\bibfnamefont
  {G.~C.}\ \bibnamefont {Gardner}}, \bibinfo {author} {\bibfnamefont {M.~J.}\
  \bibnamefont {Manfra}}, \bibinfo {author} {\bibfnamefont {E.}~\bibnamefont
  {van Nieuwenburg}}, \bibinfo {author} {\bibfnamefont {J.}~\bibnamefont
  {Danon}}, \bibinfo {author} {\bibfnamefont {A.}~\bibnamefont {Chatterjee}},\
  and\ \bibinfo {author} {\bibfnamefont {F.}~\bibnamefont {Kuemmeth}},\
  }\bibfield  {title} {\bibinfo {title} {Real-time two-axis control of a spin
  qubit},\ }\href {https://doi.org/10.1038/s41467-024-45857-0} {\bibfield
  {journal} {\bibinfo  {journal} {Nature Communications}\ }\textbf {\bibinfo
  {volume} {15}},\ \bibinfo {pages} {1676} (\bibinfo {year}
  {2024}{\natexlab{a}})}\BibitemShut {NoStop}%
\bibitem [{\citenamefont {Dumoulin~Stuyck}\ \emph {et~al.}(2024)\citenamefont
  {Dumoulin~Stuyck}, \citenamefont {Seedhouse}, \citenamefont {Serrano},
  \citenamefont {Tanttu}, \citenamefont {Gilbert}, \citenamefont {Huang},
  \citenamefont {Hudson}, \citenamefont {Itoh}, \citenamefont {Laucht},
  \citenamefont {Lim}, \citenamefont {Yang}, \citenamefont {Saraiva},\ and\
  \citenamefont {Dzurak}}]{dumoulin2024silicon}%
  \BibitemOpen
  \bibfield  {author} {\bibinfo {author} {\bibfnamefont {N.}~\bibnamefont
  {Dumoulin~Stuyck}}, \bibinfo {author} {\bibfnamefont {A.~E.}\ \bibnamefont
  {Seedhouse}}, \bibinfo {author} {\bibfnamefont {S.}~\bibnamefont {Serrano}},
  \bibinfo {author} {\bibfnamefont {T.}~\bibnamefont {Tanttu}}, \bibinfo
  {author} {\bibfnamefont {W.}~\bibnamefont {Gilbert}}, \bibinfo {author}
  {\bibfnamefont {J.~Y.}\ \bibnamefont {Huang}}, \bibinfo {author}
  {\bibfnamefont {F.}~\bibnamefont {Hudson}}, \bibinfo {author} {\bibfnamefont
  {K.~M.}\ \bibnamefont {Itoh}}, \bibinfo {author} {\bibfnamefont
  {A.}~\bibnamefont {Laucht}}, \bibinfo {author} {\bibfnamefont {W.~H.}\
  \bibnamefont {Lim}}, \bibinfo {author} {\bibfnamefont {C.~H.}\ \bibnamefont
  {Yang}}, \bibinfo {author} {\bibfnamefont {A.}~\bibnamefont {Saraiva}},\ and\
  \bibinfo {author} {\bibfnamefont {A.~S.}\ \bibnamefont {Dzurak}},\ }\bibfield
   {title} {\bibinfo {title} {Silicon spin qubit noise characterization using
  real-time feedback protocols and wavelet analysis},\ }\href
  {https://doi.org/10.1063/5.0179958} {\bibfield  {journal} {\bibinfo
  {journal} {Applied Physics Letters}\ }\textbf {\bibinfo {volume} {124}},\
  \bibinfo {pages} {114003} (\bibinfo {year} {2024})}\BibitemShut {NoStop}%
\bibitem [{\citenamefont {Vora}\ \emph {et~al.}(2024)\citenamefont {Vora},
  \citenamefont {Xu}, \citenamefont {Hashim}, \citenamefont {Fruitwala},
  \citenamefont {Nguyen}, \citenamefont {Liao}, \citenamefont {Balewski},
  \citenamefont {Rajagopala}, \citenamefont {Nowrouzi}, \citenamefont {Ji}
  \emph {et~al.}}]{vora2024ml}%
  \BibitemOpen
  \bibfield  {author} {\bibinfo {author} {\bibfnamefont {N.~R.}\ \bibnamefont
  {Vora}}, \bibinfo {author} {\bibfnamefont {Y.}~\bibnamefont {Xu}}, \bibinfo
  {author} {\bibfnamefont {A.}~\bibnamefont {Hashim}}, \bibinfo {author}
  {\bibfnamefont {N.}~\bibnamefont {Fruitwala}}, \bibinfo {author}
  {\bibfnamefont {H.~N.}\ \bibnamefont {Nguyen}}, \bibinfo {author}
  {\bibfnamefont {H.}~\bibnamefont {Liao}}, \bibinfo {author} {\bibfnamefont
  {J.}~\bibnamefont {Balewski}}, \bibinfo {author} {\bibfnamefont
  {A.}~\bibnamefont {Rajagopala}}, \bibinfo {author} {\bibfnamefont
  {K.}~\bibnamefont {Nowrouzi}}, \bibinfo {author} {\bibfnamefont
  {Q.}~\bibnamefont {Ji}}, \emph {et~al.},\ }\bibfield  {title} {\bibinfo
  {title} {{ML}-powered {FPGA}-based real-time quantum state discrimination
  enabling mid-circuit measurements}\ }\href
  {https://doi.org/10.48550/arXiv.2406.18807} {10.48550/arXiv.2406.18807}
  (\bibinfo {year} {2024})\BibitemShut {NoStop}%
\bibitem [{\citenamefont {Berritta}\ \emph
  {et~al.}(2024{\natexlab{b}})\citenamefont {Berritta}, \citenamefont
  {Krzywda}, \citenamefont {Benestad}, \citenamefont {van~der Heijden},
  \citenamefont {Fedele}, \citenamefont {Fallahi}, \citenamefont {Gardner},
  \citenamefont {Manfra}, \citenamefont {van Nieuwenburg}, \citenamefont
  {Danon}, \citenamefont {Chatterjee},\ and\ \citenamefont
  {Kuemmeth}}]{Berritta2024b}%
  \BibitemOpen
  \bibfield  {author} {\bibinfo {author} {\bibfnamefont {F.}~\bibnamefont
  {Berritta}}, \bibinfo {author} {\bibfnamefont {J.~A.}\ \bibnamefont
  {Krzywda}}, \bibinfo {author} {\bibfnamefont {J.}~\bibnamefont {Benestad}},
  \bibinfo {author} {\bibfnamefont {J.}~\bibnamefont {van~der Heijden}},
  \bibinfo {author} {\bibfnamefont {F.}~\bibnamefont {Fedele}}, \bibinfo
  {author} {\bibfnamefont {S.}~\bibnamefont {Fallahi}}, \bibinfo {author}
  {\bibfnamefont {G.~C.}\ \bibnamefont {Gardner}}, \bibinfo {author}
  {\bibfnamefont {M.~J.}\ \bibnamefont {Manfra}}, \bibinfo {author}
  {\bibfnamefont {E.}~\bibnamefont {van Nieuwenburg}}, \bibinfo {author}
  {\bibfnamefont {J.}~\bibnamefont {Danon}}, \bibinfo {author} {\bibfnamefont
  {A.}~\bibnamefont {Chatterjee}},\ and\ \bibinfo {author} {\bibfnamefont
  {F.}~\bibnamefont {Kuemmeth}},\ }\bibfield  {title} {\bibinfo {title}
  {Physics-informed tracking of qubit fluctuations},\ }\href
  {https://doi.org/10.1103/PhysRevApplied.22.014033} {\bibfield  {journal}
  {\bibinfo  {journal} {Physical Review Applied}\ }\textbf {\bibinfo {volume}
  {22}},\ \bibinfo {pages} {014033} (\bibinfo {year}
  {2024}{\natexlab{b}})}\BibitemShut {NoStop}%
\bibitem [{\citenamefont {Park}\ \emph {et~al.}(2025)\citenamefont {Park},
  \citenamefont {Jang}, \citenamefont {Sohn}, \citenamefont {Yun},
  \citenamefont {Song}, \citenamefont {Kang}, \citenamefont {Stehouwer},
  \citenamefont {Esposti}, \citenamefont {Scappucci},\ and\ \citenamefont
  {Kim}}]{Park2025}%
  \BibitemOpen
  \bibfield  {author} {\bibinfo {author} {\bibfnamefont {J.}~\bibnamefont
  {Park}}, \bibinfo {author} {\bibfnamefont {H.}~\bibnamefont {Jang}}, \bibinfo
  {author} {\bibfnamefont {H.}~\bibnamefont {Sohn}}, \bibinfo {author}
  {\bibfnamefont {J.}~\bibnamefont {Yun}}, \bibinfo {author} {\bibfnamefont
  {Y.}~\bibnamefont {Song}}, \bibinfo {author} {\bibfnamefont {B.}~\bibnamefont
  {Kang}}, \bibinfo {author} {\bibfnamefont {L.~E.}\ \bibnamefont {Stehouwer}},
  \bibinfo {author} {\bibfnamefont {D.~D.}\ \bibnamefont {Esposti}}, \bibinfo
  {author} {\bibfnamefont {G.}~\bibnamefont {Scappucci}},\ and\ \bibinfo
  {author} {\bibfnamefont {D.}~\bibnamefont {Kim}},\ }\bibfield  {title}
  {\bibinfo {title} {Passive and active suppression of transduced noise in
  silicon spin qubits},\ }\href
  {https://doi.org/https://doi.org/10.1038/s41467-024-55338-z} {\bibfield
  {journal} {\bibinfo  {journal} {Nature Communications}\ }\textbf {\bibinfo
  {volume} {16}},\ \bibinfo {pages} {78} (\bibinfo {year} {2025})}\BibitemShut
  {NoStop}%
\bibitem [{\citenamefont {Kimmel}\ \emph {et~al.}(2015)\citenamefont {Kimmel},
  \citenamefont {Low},\ and\ \citenamefont {Yoder}}]{Kimmel2015}%
  \BibitemOpen
  \bibfield  {author} {\bibinfo {author} {\bibfnamefont {S.}~\bibnamefont
  {Kimmel}}, \bibinfo {author} {\bibfnamefont {G.~H.}\ \bibnamefont {Low}},\
  and\ \bibinfo {author} {\bibfnamefont {T.~J.}\ \bibnamefont {Yoder}},\
  }\bibfield  {title} {\bibinfo {title} {Robust calibration of a universal
  single-qubit gate set via robust phase estimation},\ }\href
  {https://doi.org/https://doi.org/10.1103/PhysRevA.92.062315} {\bibfield
  {journal} {\bibinfo  {journal} {Physical Review A}\ }\textbf {\bibinfo
  {volume} {92}},\ \bibinfo {pages} {062315} (\bibinfo {year}
  {2015})}\BibitemShut {NoStop}%
\bibitem [{\citenamefont {McMichael}\ and\ \citenamefont
  {Blakley}(2022)}]{mcmichael2022}%
  \BibitemOpen
  \bibfield  {author} {\bibinfo {author} {\bibfnamefont {R.~D.}\ \bibnamefont
  {McMichael}}\ and\ \bibinfo {author} {\bibfnamefont {S.~M.}\ \bibnamefont
  {Blakley}},\ }\bibfield  {title} {\bibinfo {title} {Simplified algorithms for
  adaptive experiment design in parameter estimation},\ }\href
  {https://doi.org/10.1103/PhysRevApplied.18.054001} {\bibfield  {journal}
  {\bibinfo  {journal} {Physical review applied}\ }\textbf {\bibinfo {volume}
  {18}},\ \bibinfo {pages} {054001} (\bibinfo {year} {2022})}\BibitemShut
  {NoStop}%
\bibitem [{\citenamefont {Hurant}\ \emph {et~al.}(2024)\citenamefont {Hurant},
  \citenamefont {Sun}, \citenamefont {Jia}, \citenamefont {Kim},\ and\
  \citenamefont {Brown}}]{Hurant2024}%
  \BibitemOpen
  \bibfield  {author} {\bibinfo {author} {\bibfnamefont {T.}~\bibnamefont
  {Hurant}}, \bibinfo {author} {\bibfnamefont {K.}~\bibnamefont {Sun}},
  \bibinfo {author} {\bibfnamefont {Z.}~\bibnamefont {Jia}}, \bibinfo {author}
  {\bibfnamefont {J.}~\bibnamefont {Kim}},\ and\ \bibinfo {author}
  {\bibfnamefont {K.~R.}\ \bibnamefont {Brown}},\ }\bibfield  {title} {\bibinfo
  {title} {Few-shot, robust calibration of single qubit gates using {Bayesian}
  robust phase estimation}\ }\href {https://doi.org/10.48550/arXiv.2407.18339}
  {10.48550/arXiv.2407.18339} (\bibinfo {year} {2024})\BibitemShut {NoStop}%
\bibitem [{\citenamefont {de~Neeve}\ \emph {et~al.}(2024)\citenamefont
  {de~Neeve}, \citenamefont {Lebedev}, \citenamefont {Negnevitsky},\ and\
  \citenamefont {Home}}]{de2024time}%
  \BibitemOpen
  \bibfield  {author} {\bibinfo {author} {\bibfnamefont {B.}~\bibnamefont
  {de~Neeve}}, \bibinfo {author} {\bibfnamefont {A.~V.}\ \bibnamefont
  {Lebedev}}, \bibinfo {author} {\bibfnamefont {V.}~\bibnamefont
  {Negnevitsky}},\ and\ \bibinfo {author} {\bibfnamefont {J.~P.}\ \bibnamefont
  {Home}},\ }\bibfield  {title} {\bibinfo {title} {Time-adaptive phase
  estimation}\ }\href {https://doi.org/10.48550/arXiv.2405.08930}
  {10.48550/arXiv.2405.08930} (\bibinfo {year} {2024})\BibitemShut {NoStop}%
\bibitem [{\citenamefont {Krantz}\ \emph {et~al.}(2019)\citenamefont {Krantz},
  \citenamefont {Kjaergaard}, \citenamefont {Yan}, \citenamefont {Orlando},
  \citenamefont {Gustavsson},\ and\ \citenamefont {Oliver}}]{Krantz2019}%
  \BibitemOpen
  \bibfield  {author} {\bibinfo {author} {\bibfnamefont {P.}~\bibnamefont
  {Krantz}}, \bibinfo {author} {\bibfnamefont {M.}~\bibnamefont {Kjaergaard}},
  \bibinfo {author} {\bibfnamefont {F.}~\bibnamefont {Yan}}, \bibinfo {author}
  {\bibfnamefont {T.~P.}\ \bibnamefont {Orlando}}, \bibinfo {author}
  {\bibfnamefont {S.}~\bibnamefont {Gustavsson}},\ and\ \bibinfo {author}
  {\bibfnamefont {W.~D.}\ \bibnamefont {Oliver}},\ }\bibfield  {title}
  {\bibinfo {title} {A quantum engineer's guide to superconducting qubits},\
  }\href {https://doi.org/10.1063/1.5089550} {\bibfield  {journal} {\bibinfo
  {journal} {Applied Physics Reviews}\ }\textbf {\bibinfo {volume} {6}},\
  \bibinfo {pages} {021318} (\bibinfo {year} {2019})}\BibitemShut {NoStop}%
\bibitem [{\citenamefont {Blais}\ \emph {et~al.}(2021)\citenamefont {Blais},
  \citenamefont {Grimsmo}, \citenamefont {Girvin},\ and\ \citenamefont
  {Wallraff}}]{Blais2021}%
  \BibitemOpen
  \bibfield  {author} {\bibinfo {author} {\bibfnamefont {A.}~\bibnamefont
  {Blais}}, \bibinfo {author} {\bibfnamefont {A.~L.}\ \bibnamefont {Grimsmo}},
  \bibinfo {author} {\bibfnamefont {S.~M.}\ \bibnamefont {Girvin}},\ and\
  \bibinfo {author} {\bibfnamefont {A.}~\bibnamefont {Wallraff}},\ }\bibfield
  {title} {\bibinfo {title} {Circuit quantum electrodynamics},\ }\href
  {https://doi.org/https://doi.org/10.1103/RevModPhys.93.025005} {\bibfield
  {journal} {\bibinfo  {journal} {Reviews of Modern Physics}\ }\textbf
  {\bibinfo {volume} {93}},\ \bibinfo {pages} {025005} (\bibinfo {year}
  {2021})}\BibitemShut {NoStop}%
\bibitem [{\citenamefont {Hertzberg}\ \emph {et~al.}(2021)\citenamefont
  {Hertzberg}, \citenamefont {Zhang}, \citenamefont {Rosenblatt}, \citenamefont
  {Magesan}, \citenamefont {Smolin}, \citenamefont {Yau}, \citenamefont
  {Adiga}, \citenamefont {Sandberg}, \citenamefont {Brink}, \citenamefont
  {Chow} \emph {et~al.}}]{hertzberg2021laser}%
  \BibitemOpen
  \bibfield  {author} {\bibinfo {author} {\bibfnamefont {J.~B.}\ \bibnamefont
  {Hertzberg}}, \bibinfo {author} {\bibfnamefont {E.~J.}\ \bibnamefont
  {Zhang}}, \bibinfo {author} {\bibfnamefont {S.}~\bibnamefont {Rosenblatt}},
  \bibinfo {author} {\bibfnamefont {E.}~\bibnamefont {Magesan}}, \bibinfo
  {author} {\bibfnamefont {J.~A.}\ \bibnamefont {Smolin}}, \bibinfo {author}
  {\bibfnamefont {J.-B.}\ \bibnamefont {Yau}}, \bibinfo {author} {\bibfnamefont
  {V.~P.}\ \bibnamefont {Adiga}}, \bibinfo {author} {\bibfnamefont
  {M.}~\bibnamefont {Sandberg}}, \bibinfo {author} {\bibfnamefont
  {M.}~\bibnamefont {Brink}}, \bibinfo {author} {\bibfnamefont {J.~M.}\
  \bibnamefont {Chow}}, \emph {et~al.},\ }\bibfield  {title} {\bibinfo {title}
  {Laser-annealing {Josephson} junctions for yielding scaled-up superconducting
  quantum processors},\ }\href {https://doi.org/10.1038/s41534-021-00464-5}
  {\bibfield  {journal} {\bibinfo  {journal} {npj Quantum Information}\
  }\textbf {\bibinfo {volume} {7}},\ \bibinfo {pages} {129} (\bibinfo {year}
  {2021})}\BibitemShut {NoStop}%
\bibitem [{\citenamefont {Berke}\ \emph {et~al.}(2022)\citenamefont {Berke},
  \citenamefont {Varvelis}, \citenamefont {Trebst}, \citenamefont {Altland},\
  and\ \citenamefont {DiVincenzo}}]{berke2022transmon}%
  \BibitemOpen
  \bibfield  {author} {\bibinfo {author} {\bibfnamefont {C.}~\bibnamefont
  {Berke}}, \bibinfo {author} {\bibfnamefont {E.}~\bibnamefont {Varvelis}},
  \bibinfo {author} {\bibfnamefont {S.}~\bibnamefont {Trebst}}, \bibinfo
  {author} {\bibfnamefont {A.}~\bibnamefont {Altland}},\ and\ \bibinfo {author}
  {\bibfnamefont {D.~P.}\ \bibnamefont {DiVincenzo}},\ }\bibfield  {title}
  {\bibinfo {title} {Transmon platform for quantum computing challenged by
  chaotic fluctuations},\ }\href {https://doi.org/10.1038/s41467-022-29940-y}
  {\bibfield  {journal} {\bibinfo  {journal} {Nature communications}\ }\textbf
  {\bibinfo {volume} {13}},\ \bibinfo {pages} {2495} (\bibinfo {year}
  {2022})}\BibitemShut {NoStop}%
\bibitem [{\citenamefont {Vepsäläinen}\ \emph {et~al.}(2022)\citenamefont
  {Vepsäläinen}, \citenamefont {Winik}, \citenamefont {Karamlou},
  \citenamefont {Braumüller}, \citenamefont {Paolo}, \citenamefont {Sung},
  \citenamefont {Kannan}, \citenamefont {Kjaergaard}, \citenamefont {Kim},
  \citenamefont {Melville}, \citenamefont {Niedzielski}, \citenamefont {Yoder},
  \citenamefont {Gustavsson},\ and\ \citenamefont
  {Oliver}}]{Vepsaelaeinen2022}%
  \BibitemOpen
  \bibfield  {author} {\bibinfo {author} {\bibfnamefont {A.}~\bibnamefont
  {Vepsäläinen}}, \bibinfo {author} {\bibfnamefont {R.}~\bibnamefont
  {Winik}}, \bibinfo {author} {\bibfnamefont {A.~H.}\ \bibnamefont {Karamlou}},
  \bibinfo {author} {\bibfnamefont {J.}~\bibnamefont {Braumüller}}, \bibinfo
  {author} {\bibfnamefont {A.~D.}\ \bibnamefont {Paolo}}, \bibinfo {author}
  {\bibfnamefont {Y.}~\bibnamefont {Sung}}, \bibinfo {author} {\bibfnamefont
  {B.}~\bibnamefont {Kannan}}, \bibinfo {author} {\bibfnamefont
  {M.}~\bibnamefont {Kjaergaard}}, \bibinfo {author} {\bibfnamefont {D.~K.}\
  \bibnamefont {Kim}}, \bibinfo {author} {\bibfnamefont {A.~J.}\ \bibnamefont
  {Melville}}, \bibinfo {author} {\bibfnamefont {B.~M.}\ \bibnamefont
  {Niedzielski}}, \bibinfo {author} {\bibfnamefont {J.~L.}\ \bibnamefont
  {Yoder}}, \bibinfo {author} {\bibfnamefont {S.}~\bibnamefont {Gustavsson}},\
  and\ \bibinfo {author} {\bibfnamefont {W.~D.}\ \bibnamefont {Oliver}},\
  }\bibfield  {title} {\bibinfo {title} {Improving qubit coherence using
  closed-loop feedback},\ }\href {https://doi.org/10.1038/s41467-022-29287-4}
  {\bibfield  {journal} {\bibinfo  {journal} {Nature Communications}\ }\textbf
  {\bibinfo {volume} {13}},\ \bibinfo {pages} {1932} (\bibinfo {year}
  {2022})}\BibitemShut {NoStop}%
\bibitem [{\citenamefont {Ferrie}\ \emph {et~al.}(2013)\citenamefont {Ferrie},
  \citenamefont {Granade},\ and\ \citenamefont {Cory}}]{Ferrie2013}%
  \BibitemOpen
  \bibfield  {author} {\bibinfo {author} {\bibfnamefont {C.}~\bibnamefont
  {Ferrie}}, \bibinfo {author} {\bibfnamefont {C.~E.}\ \bibnamefont
  {Granade}},\ and\ \bibinfo {author} {\bibfnamefont {D.~G.}\ \bibnamefont
  {Cory}},\ }\bibfield  {title} {\bibinfo {title} {How to best sample a
  periodic probability distribution, or on the accuracy of {Hamiltonian}
  finding strategies},\ }\href
  {https://doi.org/https://doi.org/10.1007/s11128-012-0407-6} {\bibfield
  {journal} {\bibinfo  {journal} {Quantum Information Processing}\ }\textbf
  {\bibinfo {volume} {12}},\ \bibinfo {pages} {611} (\bibinfo {year}
  {2013})}\BibitemShut {NoStop}%
\bibitem [{Note1()}]{Note1}%
  \BibitemOpen
  \bibinfo {note} {The commonly used term \protect \emph {non-Markovian noise}
  refers to the non-unitary evolution of a quantum system that cannot be
  modeled by the Markovian master equation. Such non-Markovian decoherence
  arises from the environment's finite memory, which introduces correlations
  between consecutive measurement outcomes.}\BibitemShut {Stop}%
\bibitem [{\citenamefont {Hakoshima}\ \emph {et~al.}(2021)\citenamefont
  {Hakoshima}, \citenamefont {Matsuzaki},\ and\ \citenamefont
  {Endo}}]{hakoshima2021relationship}%
  \BibitemOpen
  \bibfield  {author} {\bibinfo {author} {\bibfnamefont {H.}~\bibnamefont
  {Hakoshima}}, \bibinfo {author} {\bibfnamefont {Y.}~\bibnamefont
  {Matsuzaki}},\ and\ \bibinfo {author} {\bibfnamefont {S.}~\bibnamefont
  {Endo}},\ }\bibfield  {title} {\bibinfo {title} {Relationship between costs
  for quantum error mitigation and {non-Markovian} measures},\ }\href
  {https://doi.org/10.1103/PhysRevA.103.012611} {\bibfield  {journal} {\bibinfo
   {journal} {Physical Review A}\ }\textbf {\bibinfo {volume} {103}},\ \bibinfo
  {pages} {012611} (\bibinfo {year} {2021})}\BibitemShut {NoStop}%
\bibitem [{\citenamefont {Kam}\ \emph {et~al.}(2024)\citenamefont {Kam},
  \citenamefont {Gicev}, \citenamefont {Modi}, \citenamefont {Southwell},\ and\
  \citenamefont {Usman}}]{kam2024detrimental}%
  \BibitemOpen
  \bibfield  {author} {\bibinfo {author} {\bibfnamefont {J.~F.}\ \bibnamefont
  {Kam}}, \bibinfo {author} {\bibfnamefont {S.}~\bibnamefont {Gicev}}, \bibinfo
  {author} {\bibfnamefont {K.}~\bibnamefont {Modi}}, \bibinfo {author}
  {\bibfnamefont {A.}~\bibnamefont {Southwell}},\ and\ \bibinfo {author}
  {\bibfnamefont {M.}~\bibnamefont {Usman}},\ }\bibfield  {title} {\bibinfo
  {title} {Detrimental {non-Markovian} errors for surface code memory}\ }\href
  {https://doi.org/10.48550/arXiv.2410.23779} {10.48550/arXiv.2410.23779}
  (\bibinfo {year} {2024})\BibitemShut {NoStop}%
\bibitem [{\citenamefont {Viola}\ \emph {et~al.}(1999)\citenamefont {Viola},
  \citenamefont {Knill},\ and\ \citenamefont {Lloyd}}]{Viola1999}%
  \BibitemOpen
  \bibfield  {author} {\bibinfo {author} {\bibfnamefont {L.}~\bibnamefont
  {Viola}}, \bibinfo {author} {\bibfnamefont {E.}~\bibnamefont {Knill}},\ and\
  \bibinfo {author} {\bibfnamefont {S.}~\bibnamefont {Lloyd}},\ }\bibfield
  {title} {\bibinfo {title} {Dynamical decoupling of open quantum systems},\
  }\href {https://doi.org/10.1103/physrevlett.82.2417} {\bibfield  {journal}
  {\bibinfo  {journal} {Physical Review Letters}\ }\textbf {\bibinfo {volume}
  {82}},\ \bibinfo {pages} {2417} (\bibinfo {year} {1999})}\BibitemShut
  {NoStop}%
\bibitem [{\citenamefont {Sza{\'{n}}kowski}\ \emph {et~al.}(2017)\citenamefont
  {Sza{\'{n}}kowski}, \citenamefont {Ramon}, \citenamefont {Krzywda},
  \citenamefont {Kwiatkowski},\ and\ \citenamefont
  {Cywi{\'{n}}ski}}]{Szankowski2017}%
  \BibitemOpen
  \bibfield  {author} {\bibinfo {author} {\bibfnamefont {P.}~\bibnamefont
  {Sza{\'{n}}kowski}}, \bibinfo {author} {\bibfnamefont {G.}~\bibnamefont
  {Ramon}}, \bibinfo {author} {\bibfnamefont {J.}~\bibnamefont {Krzywda}},
  \bibinfo {author} {\bibfnamefont {D.}~\bibnamefont {Kwiatkowski}},\ and\
  \bibinfo {author} {\bibfnamefont {{\L}.}~\bibnamefont {Cywi{\'{n}}ski}},\
  }\bibfield  {title} {\bibinfo {title} {Environmental noise spectroscopy with
  qubits subjected to dynamical decoupling},\ }\href
  {https://doi.org/10.1088/1361-648x/aa7648} {\bibfield  {journal} {\bibinfo
  {journal} {Journal of Physics: Condensed Matter}\ }\textbf {\bibinfo {volume}
  {29}},\ \bibinfo {pages} {333001} (\bibinfo {year} {2017})}\BibitemShut
  {NoStop}%
\bibitem [{\citenamefont {Pataki}\ \emph {et~al.}(2024)\citenamefont {Pataki},
  \citenamefont {M{\'a}rton}, \citenamefont {Asb{\'o}th},\ and\ \citenamefont
  {P{\'a}lyi}}]{pataki2024coherent}%
  \BibitemOpen
  \bibfield  {author} {\bibinfo {author} {\bibfnamefont {D.}~\bibnamefont
  {Pataki}}, \bibinfo {author} {\bibfnamefont {{\'A}.}~\bibnamefont
  {M{\'a}rton}}, \bibinfo {author} {\bibfnamefont {J.~K.}\ \bibnamefont
  {Asb{\'o}th}},\ and\ \bibinfo {author} {\bibfnamefont {A.}~\bibnamefont
  {P{\'a}lyi}},\ }\bibfield  {title} {\bibinfo {title} {Coherent errors in
  stabilizer codes caused by quasistatic phase damping},\ }\href
  {https://doi.org/https://doi.org/10.1103/PhysRevA.110.012417} {\bibfield
  {journal} {\bibinfo  {journal} {Physical Review A}\ }\textbf {\bibinfo
  {volume} {110}},\ \bibinfo {pages} {012417} (\bibinfo {year}
  {2024})}\BibitemShut {NoStop}%
\bibitem [{\citenamefont {Nielsen}\ \emph {et~al.}(2021)\citenamefont
  {Nielsen}, \citenamefont {Gamble}, \citenamefont {Rudinger}, \citenamefont
  {Scholten}, \citenamefont {Young},\ and\ \citenamefont
  {Blume-Kohout}}]{Nielsen2021}%
  \BibitemOpen
  \bibfield  {author} {\bibinfo {author} {\bibfnamefont {E.}~\bibnamefont
  {Nielsen}}, \bibinfo {author} {\bibfnamefont {J.~K.}\ \bibnamefont {Gamble}},
  \bibinfo {author} {\bibfnamefont {K.}~\bibnamefont {Rudinger}}, \bibinfo
  {author} {\bibfnamefont {T.}~\bibnamefont {Scholten}}, \bibinfo {author}
  {\bibfnamefont {K.}~\bibnamefont {Young}},\ and\ \bibinfo {author}
  {\bibfnamefont {R.}~\bibnamefont {Blume-Kohout}},\ }\bibfield  {title}
  {\bibinfo {title} {Gate set tomography},\ }\href
  {https://doi.org/10.22331/q-2021-10-05-557} {\bibfield  {journal} {\bibinfo
  {journal} {Quantum}\ }\textbf {\bibinfo {volume} {5}},\ \bibinfo {pages}
  {557} (\bibinfo {year} {2021})}\BibitemShut {NoStop}%
\bibitem [{\citenamefont {Knill}\ \emph {et~al.}(2008)\citenamefont {Knill},
  \citenamefont {Leibfried}, \citenamefont {Reichle}, \citenamefont {Britton},
  \citenamefont {Blakestad}, \citenamefont {Jost}, \citenamefont {Langer},
  \citenamefont {Ozeri}, \citenamefont {Seidelin},\ and\ \citenamefont
  {Wineland}}]{Knill2008}%
  \BibitemOpen
  \bibfield  {author} {\bibinfo {author} {\bibfnamefont {E.}~\bibnamefont
  {Knill}}, \bibinfo {author} {\bibfnamefont {D.}~\bibnamefont {Leibfried}},
  \bibinfo {author} {\bibfnamefont {R.}~\bibnamefont {Reichle}}, \bibinfo
  {author} {\bibfnamefont {J.}~\bibnamefont {Britton}}, \bibinfo {author}
  {\bibfnamefont {R.~B.}\ \bibnamefont {Blakestad}}, \bibinfo {author}
  {\bibfnamefont {J.~D.}\ \bibnamefont {Jost}}, \bibinfo {author}
  {\bibfnamefont {C.}~\bibnamefont {Langer}}, \bibinfo {author} {\bibfnamefont
  {R.}~\bibnamefont {Ozeri}}, \bibinfo {author} {\bibfnamefont
  {S.}~\bibnamefont {Seidelin}},\ and\ \bibinfo {author} {\bibfnamefont
  {D.~J.}\ \bibnamefont {Wineland}},\ }\bibfield  {title} {\bibinfo {title}
  {Randomized benchmarking of quantum gates},\ }\href
  {https://doi.org/10.1103/physreva.77.012307} {\bibfield  {journal} {\bibinfo
  {journal} {Physical Review A}\ }\textbf {\bibinfo {volume} {77}},\ \bibinfo
  {pages} {012307} (\bibinfo {year} {2008})}\BibitemShut {NoStop}%
\bibitem [{\citenamefont {Sung}\ \emph {et~al.}(2021)\citenamefont {Sung},
  \citenamefont {Ding}, \citenamefont {Braumüller}, \citenamefont
  {Vepsäläinen}, \citenamefont {Kannan}, \citenamefont {Kjaergaard},
  \citenamefont {Greene}, \citenamefont {Samach}, \citenamefont {McNally},
  \citenamefont {Kim}, \citenamefont {Melville}, \citenamefont {Niedzielski},
  \citenamefont {Schwartz}, \citenamefont {Yoder}, \citenamefont {Orlando},
  \citenamefont {Gustavsson},\ and\ \citenamefont {Oliver}}]{Sung2021}%
  \BibitemOpen
  \bibfield  {author} {\bibinfo {author} {\bibfnamefont {Y.}~\bibnamefont
  {Sung}}, \bibinfo {author} {\bibfnamefont {L.}~\bibnamefont {Ding}}, \bibinfo
  {author} {\bibfnamefont {J.}~\bibnamefont {Braumüller}}, \bibinfo {author}
  {\bibfnamefont {A.}~\bibnamefont {Vepsäläinen}}, \bibinfo {author}
  {\bibfnamefont {B.}~\bibnamefont {Kannan}}, \bibinfo {author} {\bibfnamefont
  {M.}~\bibnamefont {Kjaergaard}}, \bibinfo {author} {\bibfnamefont
  {A.}~\bibnamefont {Greene}}, \bibinfo {author} {\bibfnamefont {G.~O.}\
  \bibnamefont {Samach}}, \bibinfo {author} {\bibfnamefont {C.}~\bibnamefont
  {McNally}}, \bibinfo {author} {\bibfnamefont {D.}~\bibnamefont {Kim}},
  \bibinfo {author} {\bibfnamefont {A.}~\bibnamefont {Melville}}, \bibinfo
  {author} {\bibfnamefont {B.~M.}\ \bibnamefont {Niedzielski}}, \bibinfo
  {author} {\bibfnamefont {M.~E.}\ \bibnamefont {Schwartz}}, \bibinfo {author}
  {\bibfnamefont {J.~L.}\ \bibnamefont {Yoder}}, \bibinfo {author}
  {\bibfnamefont {T.~P.}\ \bibnamefont {Orlando}}, \bibinfo {author}
  {\bibfnamefont {S.}~\bibnamefont {Gustavsson}},\ and\ \bibinfo {author}
  {\bibfnamefont {W.~D.}\ \bibnamefont {Oliver}},\ }\bibfield  {title}
  {\bibinfo {title} {Realization of high-fidelity {CZ} and {ZZ} -free {iSWAP}
  gates with a tunable coupler},\ }\href
  {https://doi.org/10.1103/physrevx.11.021058} {\bibfield  {journal} {\bibinfo
  {journal} {Physical Review X}\ }\textbf {\bibinfo {volume} {11}},\ \bibinfo
  {pages} {021058} (\bibinfo {year} {2021})}\BibitemShut {NoStop}%
\bibitem [{sup()}]{supplementary}%
  \BibitemOpen
  \href@noop {} {}\bibinfo {note} {See Supplemental Material, which includes
  Refs.~\cite{rojas2024,paquelet2023reducing,Magesan2012,huber_robust_1981}.}\BibitemShut
  {Stop}%
\bibitem [{\citenamefont {Bonato}\ \emph {et~al.}(2015)\citenamefont {Bonato},
  \citenamefont {Blok}, \citenamefont {Dinani}, \citenamefont {Berry},
  \citenamefont {Markham}, \citenamefont {Twitchen},\ and\ \citenamefont
  {Hanson}}]{Bonato2015}%
  \BibitemOpen
  \bibfield  {author} {\bibinfo {author} {\bibfnamefont {C.}~\bibnamefont
  {Bonato}}, \bibinfo {author} {\bibfnamefont {M.~S.}\ \bibnamefont {Blok}},
  \bibinfo {author} {\bibfnamefont {H.~T.}\ \bibnamefont {Dinani}}, \bibinfo
  {author} {\bibfnamefont {D.~W.}\ \bibnamefont {Berry}}, \bibinfo {author}
  {\bibfnamefont {M.~L.}\ \bibnamefont {Markham}}, \bibinfo {author}
  {\bibfnamefont {D.~J.}\ \bibnamefont {Twitchen}},\ and\ \bibinfo {author}
  {\bibfnamefont {R.}~\bibnamefont {Hanson}},\ }\bibfield  {title} {\bibinfo
  {title} {Optimized quantum sensing with a single electron spin using
  real-time adaptive measurements},\ }\href
  {https://doi.org/10.1038/nnano.2015.261} {\bibfield  {journal} {\bibinfo
  {journal} {Nature Nanotechnology}\ }\textbf {\bibinfo {volume} {11}},\
  \bibinfo {pages} {247} (\bibinfo {year} {2015})}\BibitemShut {NoStop}%
\bibitem [{\citenamefont {Joas}\ \emph {et~al.}(2021)\citenamefont {Joas},
  \citenamefont {Schmitt}, \citenamefont {Santagati}, \citenamefont {Gentile},
  \citenamefont {Bonato}, \citenamefont {Laing}, \citenamefont {McGuinness},\
  and\ \citenamefont {Jelezko}}]{joas2021}%
  \BibitemOpen
  \bibfield  {author} {\bibinfo {author} {\bibfnamefont {T.}~\bibnamefont
  {Joas}}, \bibinfo {author} {\bibfnamefont {S.}~\bibnamefont {Schmitt}},
  \bibinfo {author} {\bibfnamefont {R.}~\bibnamefont {Santagati}}, \bibinfo
  {author} {\bibfnamefont {A.~A.}\ \bibnamefont {Gentile}}, \bibinfo {author}
  {\bibfnamefont {C.}~\bibnamefont {Bonato}}, \bibinfo {author} {\bibfnamefont
  {A.}~\bibnamefont {Laing}}, \bibinfo {author} {\bibfnamefont {L.~P.}\
  \bibnamefont {McGuinness}},\ and\ \bibinfo {author} {\bibfnamefont
  {F.}~\bibnamefont {Jelezko}},\ }\bibfield  {title} {\bibinfo {title} {Online
  adaptive quantum characterization of a nuclear spin},\ }\href
  {https://doi.org/10.1038/s41534-021-00389-z} {\bibfield  {journal} {\bibinfo
  {journal} {npj Quantum Information}\ }\textbf {\bibinfo {volume} {7}},\
  \bibinfo {pages} {56} (\bibinfo {year} {2021})}\BibitemShut {NoStop}%
\bibitem [{\citenamefont {Childs}\ \emph {et~al.}(2000)\citenamefont {Childs},
  \citenamefont {Preskill},\ and\ \citenamefont {Renes}}]{childs_quantum_2000}%
  \BibitemOpen
  \bibfield  {author} {\bibinfo {author} {\bibfnamefont {A.~M.}\ \bibnamefont
  {Childs}}, \bibinfo {author} {\bibfnamefont {J.}~\bibnamefont {Preskill}},\
  and\ \bibinfo {author} {\bibfnamefont {J.}~\bibnamefont {Renes}},\ }\bibfield
   {title} {\bibinfo {title} {Quantum information and precision measurement},\
  }\href {https://doi.org/https://doi.org/10.1080/09500340008244034} {\bibfield
   {journal} {\bibinfo  {journal} {Journal of modern optics}\ }\textbf
  {\bibinfo {volume} {47}},\ \bibinfo {pages} {155} (\bibinfo {year}
  {2000})}\BibitemShut {NoStop}%
\bibitem [{\citenamefont {Said}\ \emph {et~al.}(2011)\citenamefont {Said},
  \citenamefont {Berry},\ and\ \citenamefont
  {Twamley}}]{said_nanoscale_magnetometry_2011}%
  \BibitemOpen
  \bibfield  {author} {\bibinfo {author} {\bibfnamefont {R.~S.}\ \bibnamefont
  {Said}}, \bibinfo {author} {\bibfnamefont {D.~W.}\ \bibnamefont {Berry}},\
  and\ \bibinfo {author} {\bibfnamefont {J.}~\bibnamefont {Twamley}},\
  }\bibfield  {title} {\bibinfo {title} {Nanoscale magnetometry using a
  single-spin system in diamond},\ }\href
  {https://doi.org/10.1103/PhysRevB.83.125410} {\bibfield  {journal} {\bibinfo
  {journal} {Physical Review B}\ }\textbf {\bibinfo {volume} {83}},\ \bibinfo
  {pages} {125410} (\bibinfo {year} {2011})}\BibitemShut {NoStop}%
\bibitem [{\citenamefont
  {Cappellaro}(2012)}]{cappellaro_spin-bath_narrowing_2012}%
  \BibitemOpen
  \bibfield  {author} {\bibinfo {author} {\bibfnamefont {P.}~\bibnamefont
  {Cappellaro}},\ }\bibfield  {title} {\bibinfo {title} {Spin-bath narrowing
  with adaptive parameter estimation},\ }\href
  {https://doi.org/10.1103/PhysRevA.85.030301} {\bibfield  {journal} {\bibinfo
  {journal} {Physical Review A}\ }\textbf {\bibinfo {volume} {85}},\ \bibinfo
  {pages} {030301} (\bibinfo {year} {2012})}\BibitemShut {NoStop}%
\bibitem [{\citenamefont {Nusran}\ \emph {et~al.}(2012)\citenamefont {Nusran},
  \citenamefont {Momeen},\ and\ \citenamefont
  {Dutt}}]{nusran_high-dynamic-range_2012}%
  \BibitemOpen
  \bibfield  {author} {\bibinfo {author} {\bibfnamefont {N.~M.}\ \bibnamefont
  {Nusran}}, \bibinfo {author} {\bibfnamefont {M.~U.}\ \bibnamefont {Momeen}},\
  and\ \bibinfo {author} {\bibfnamefont {M.~V.~G.}\ \bibnamefont {Dutt}},\
  }\bibfield  {title} {\bibinfo {title} {High-dynamic-range magnetometry with a
  single electronic spin in diamond},\ }\href
  {https://doi.org/10.1038/nnano.2011.225} {\bibfield  {journal} {\bibinfo
  {journal} {Nature Nanotechnology}\ }\textbf {\bibinfo {volume} {7}},\
  \bibinfo {pages} {109} (\bibinfo {year} {2012})}\BibitemShut {NoStop}%
\bibitem [{\citenamefont {Waldherr}\ \emph {et~al.}(2012)\citenamefont
  {Waldherr}, \citenamefont {Beck}, \citenamefont {Neumann}, \citenamefont
  {Said}, \citenamefont {Nitsche}, \citenamefont {Markham}, \citenamefont
  {Twitchen}, \citenamefont {Twamley}, \citenamefont {Jelezko},\ and\
  \citenamefont {Wrachtrup}}]{waldherr_high-dynamic-range_2012}%
  \BibitemOpen
  \bibfield  {author} {\bibinfo {author} {\bibfnamefont {G.}~\bibnamefont
  {Waldherr}}, \bibinfo {author} {\bibfnamefont {J.}~\bibnamefont {Beck}},
  \bibinfo {author} {\bibfnamefont {P.}~\bibnamefont {Neumann}}, \bibinfo
  {author} {\bibfnamefont {R.~S.}\ \bibnamefont {Said}}, \bibinfo {author}
  {\bibfnamefont {M.}~\bibnamefont {Nitsche}}, \bibinfo {author} {\bibfnamefont
  {M.~L.}\ \bibnamefont {Markham}}, \bibinfo {author} {\bibfnamefont {D.~J.}\
  \bibnamefont {Twitchen}}, \bibinfo {author} {\bibfnamefont {J.}~\bibnamefont
  {Twamley}}, \bibinfo {author} {\bibfnamefont {F.}~\bibnamefont {Jelezko}},\
  and\ \bibinfo {author} {\bibfnamefont {J.}~\bibnamefont {Wrachtrup}},\
  }\bibfield  {title} {\bibinfo {title} {High-dynamic-range magnetometry with a
  single nuclear spin in diamond},\ }\href
  {https://doi.org/10.1038/nnano.2011.224} {\bibfield  {journal} {\bibinfo
  {journal} {Nature Nanotechnology}\ }\textbf {\bibinfo {volume} {7}},\
  \bibinfo {pages} {105} (\bibinfo {year} {2012})}\BibitemShut {NoStop}%
\bibitem [{Note2()}]{Note2}%
  \BibitemOpen
  \bibinfo {note} {We clarify that this is not a true binary search in the
  sense of gaining exactly one bit of information per measurement. The search
  does, however, follow a binary search tree where the two options at each node
  provide the most information within the approximations we use. A true
  \protect \emph {quantum} binary search could be implemented for a
  decoherence-free qubit with ideal initialization and readout~\cite
  {childs_quantum_2000}. The focus here is on estimating noise in a physical
  qubit.}\BibitemShut {Stop}%
\bibitem [{Note3()}]{Note3}%
  \BibitemOpen
  \bibinfo {note} {In the limit $\mu /\sigma \rightarrow 0$, for $m = +1$, the
  likelihood function with zero phase (a squared cosine) has a single maximum
  at $\mu = 0$ within the 95\% credible interval of the prior distribution, and
  the corresponding posterior distribution can be approximated by a Gaussian.
  In contrast, for $m = -1$ in the same limit, the zero-phase likelihood
  function (a squared sine) exhibits two global maxima within the 95\% credible
  interval and a minimum at the maximum prior value, resulting in a bimodal
  posterior distribution that cannot be adequately approximated by a single
  Gaussian.}\BibitemShut {Stop}%
\bibitem [{\citenamefont {Benestad}\ \emph {et~al.}(2024)\citenamefont
  {Benestad}, \citenamefont {Krzywda}, \citenamefont {van Nieuwenburg},\ and\
  \citenamefont {Danon}}]{Benestad2024}%
  \BibitemOpen
  \bibfield  {author} {\bibinfo {author} {\bibfnamefont {J.}~\bibnamefont
  {Benestad}}, \bibinfo {author} {\bibfnamefont {J.~A.}\ \bibnamefont
  {Krzywda}}, \bibinfo {author} {\bibfnamefont {E.}~\bibnamefont {van
  Nieuwenburg}},\ and\ \bibinfo {author} {\bibfnamefont {J.}~\bibnamefont
  {Danon}},\ }\bibfield  {title} {\bibinfo {title} {{Efficient adaptive
  Bayesian estimation of a slowly fluctuating Overhauser field gradient}},\
  }\href {https://doi.org/10.21468/SciPostPhys.17.1.014} {\bibfield  {journal}
  {\bibinfo  {journal} {SciPost Physics}\ }\textbf {\bibinfo {volume} {17}},\
  \bibinfo {pages} {014} (\bibinfo {year} {2024})}\BibitemShut {NoStop}%
\bibitem [{Note4()}]{Note4}%
  \BibitemOpen
  \bibinfo {note} {Note that we do not claim exponential scaling with sensing
  time, which is fundamentally bounded by the Heisenberg limit $\sigma
  ^2(\varepsilon ) \sim 1/\tau ^2$. Instead, our protocol achieves exponential
  scaling with the number of single-shot measurements $N$.}\BibitemShut {Stop}%
\bibitem [{\citenamefont {McKay}\ \emph {et~al.}(2017)\citenamefont {McKay},
  \citenamefont {Wood}, \citenamefont {Sheldon}, \citenamefont {Chow},\ and\
  \citenamefont {Gambetta}}]{McKay2017}%
  \BibitemOpen
  \bibfield  {author} {\bibinfo {author} {\bibfnamefont {D.~C.}\ \bibnamefont
  {McKay}}, \bibinfo {author} {\bibfnamefont {C.~J.}\ \bibnamefont {Wood}},
  \bibinfo {author} {\bibfnamefont {S.}~\bibnamefont {Sheldon}}, \bibinfo
  {author} {\bibfnamefont {J.~M.}\ \bibnamefont {Chow}},\ and\ \bibinfo
  {author} {\bibfnamefont {J.~M.}\ \bibnamefont {Gambetta}},\ }\bibfield
  {title} {\bibinfo {title} {Efficient {Z} gates for quantum computing},\
  }\href {https://doi.org/10.1103/physreva.96.022330} {\bibfield  {journal}
  {\bibinfo  {journal} {Physical Review A}\ }\textbf {\bibinfo {volume} {96}},\
  \bibinfo {pages} {022330} (\bibinfo {year} {2017})}\BibitemShut {NoStop}%
\bibitem [{Note5()}]{Note5}%
  \BibitemOpen
  \bibinfo {note} {In the RB experiment, $\mu _0$ is reset to zero before each
  estimation sequence, which explains the increased $\sigma _0$ and $N$
  compared to the Ramsey experiment of the previous section, where we do not
  reset $\mu _0$ before each estimation sequence. We believe that further
  improvements in the single-qubit gate fidelity could be achieved by using the
  same settings as in the Ramsey experiment. Still, this work focuses on
  demonstrating the enhancement provided by FBS, rather than achieving the
  highest possible fidelity for this particular setup.}\BibitemShut {Stop}%
\bibitem [{\citenamefont {O’Malley}\ \emph {et~al.}(2015)\citenamefont
  {O’Malley}, \citenamefont {Kelly}, \citenamefont {Barends}, \citenamefont
  {Campbell}, \citenamefont {Chen}, \citenamefont {Chen}, \citenamefont
  {Chiaro}, \citenamefont {Dunsworth}, \citenamefont {Fowler}, \citenamefont
  {Hoi}, \citenamefont {Jeffrey}, \citenamefont {Megrant}, \citenamefont
  {Mutus}, \citenamefont {Neill}, \citenamefont {Quintana}, \citenamefont
  {Roushan}, \citenamefont {Sank}, \citenamefont {Vainsencher}, \citenamefont
  {Wenner}, \citenamefont {White}, \citenamefont {Korotkov}, \citenamefont
  {Cleland},\ and\ \citenamefont {Martinis}}]{Malley2015}%
  \BibitemOpen
  \bibfield  {author} {\bibinfo {author} {\bibfnamefont {P.~J.~J.}\
  \bibnamefont {O’Malley}}, \bibinfo {author} {\bibfnamefont
  {J.}~\bibnamefont {Kelly}}, \bibinfo {author} {\bibfnamefont
  {R.}~\bibnamefont {Barends}}, \bibinfo {author} {\bibfnamefont
  {B.}~\bibnamefont {Campbell}}, \bibinfo {author} {\bibfnamefont
  {Y.}~\bibnamefont {Chen}}, \bibinfo {author} {\bibfnamefont {Z.}~\bibnamefont
  {Chen}}, \bibinfo {author} {\bibfnamefont {B.}~\bibnamefont {Chiaro}},
  \bibinfo {author} {\bibfnamefont {A.}~\bibnamefont {Dunsworth}}, \bibinfo
  {author} {\bibfnamefont {A.~G.}\ \bibnamefont {Fowler}}, \bibinfo {author}
  {\bibfnamefont {I.-C.}\ \bibnamefont {Hoi}}, \bibinfo {author} {\bibfnamefont
  {E.}~\bibnamefont {Jeffrey}}, \bibinfo {author} {\bibfnamefont
  {A.}~\bibnamefont {Megrant}}, \bibinfo {author} {\bibfnamefont
  {J.}~\bibnamefont {Mutus}}, \bibinfo {author} {\bibfnamefont
  {C.}~\bibnamefont {Neill}}, \bibinfo {author} {\bibfnamefont
  {C.}~\bibnamefont {Quintana}}, \bibinfo {author} {\bibfnamefont
  {P.}~\bibnamefont {Roushan}}, \bibinfo {author} {\bibfnamefont
  {D.}~\bibnamefont {Sank}}, \bibinfo {author} {\bibfnamefont {A.}~\bibnamefont
  {Vainsencher}}, \bibinfo {author} {\bibfnamefont {J.}~\bibnamefont {Wenner}},
  \bibinfo {author} {\bibfnamefont {T.~C.}\ \bibnamefont {White}}, \bibinfo
  {author} {\bibfnamefont {A.~N.}\ \bibnamefont {Korotkov}}, \bibinfo {author}
  {\bibfnamefont {A.~N.}\ \bibnamefont {Cleland}},\ and\ \bibinfo {author}
  {\bibfnamefont {J.~M.}\ \bibnamefont {Martinis}},\ }\bibfield  {title}
  {\bibinfo {title} {Qubit metrology of ultralow phase noise using randomized
  benchmarking},\ }\href {https://doi.org/10.1103/physrevapplied.3.044009}
  {\bibfield  {journal} {\bibinfo  {journal} {Physical Review Applied}\
  }\textbf {\bibinfo {volume} {3}},\ \bibinfo {pages} {044009} (\bibinfo {year}
  {2015})}\BibitemShut {NoStop}%
\bibitem [{\citenamefont {Nielsen}\ \emph {et~al.}(2020)\citenamefont
  {Nielsen}, \citenamefont {Rudinger}, \citenamefont {Proctor}, \citenamefont
  {Russo}, \citenamefont {Young},\ and\ \citenamefont
  {Blume-Kohout}}]{Nielsen2020}%
  \BibitemOpen
  \bibfield  {author} {\bibinfo {author} {\bibfnamefont {E.}~\bibnamefont
  {Nielsen}}, \bibinfo {author} {\bibfnamefont {K.}~\bibnamefont {Rudinger}},
  \bibinfo {author} {\bibfnamefont {T.}~\bibnamefont {Proctor}}, \bibinfo
  {author} {\bibfnamefont {A.}~\bibnamefont {Russo}}, \bibinfo {author}
  {\bibfnamefont {K.}~\bibnamefont {Young}},\ and\ \bibinfo {author}
  {\bibfnamefont {R.}~\bibnamefont {Blume-Kohout}},\ }\bibfield  {title}
  {\bibinfo {title} {Probing quantum processor performance with {pyGSTi}},\
  }\href {https://doi.org/10.1088/2058-9565/ab8aa4} {\bibfield  {journal}
  {\bibinfo  {journal} {Quantum Science and Technology}\ }\textbf {\bibinfo
  {volume} {5}},\ \bibinfo {pages} {044002} (\bibinfo {year}
  {2020})}\BibitemShut {NoStop}%
\bibitem [{\citenamefont {Hashim}\ \emph {et~al.}(2024)\citenamefont {Hashim},
  \citenamefont {Nguyen}, \citenamefont {Goss}, \citenamefont {Marinelli},
  \citenamefont {Naik}, \citenamefont {Chistolini}, \citenamefont {Hines},
  \citenamefont {Marceaux}, \citenamefont {Kim}, \citenamefont {Gokhale} \emph
  {et~al.}}]{hashim2024practical}%
  \BibitemOpen
  \bibfield  {author} {\bibinfo {author} {\bibfnamefont {A.}~\bibnamefont
  {Hashim}}, \bibinfo {author} {\bibfnamefont {L.~B.}\ \bibnamefont {Nguyen}},
  \bibinfo {author} {\bibfnamefont {N.}~\bibnamefont {Goss}}, \bibinfo {author}
  {\bibfnamefont {B.}~\bibnamefont {Marinelli}}, \bibinfo {author}
  {\bibfnamefont {R.~K.}\ \bibnamefont {Naik}}, \bibinfo {author}
  {\bibfnamefont {T.}~\bibnamefont {Chistolini}}, \bibinfo {author}
  {\bibfnamefont {J.}~\bibnamefont {Hines}}, \bibinfo {author} {\bibfnamefont
  {J.}~\bibnamefont {Marceaux}}, \bibinfo {author} {\bibfnamefont
  {Y.}~\bibnamefont {Kim}}, \bibinfo {author} {\bibfnamefont {P.}~\bibnamefont
  {Gokhale}}, \emph {et~al.},\ }\bibfield  {title} {\bibinfo {title} {A
  practical introduction to benchmarking and characterization of quantum
  computers}\ }\href {https://doi.org/10.48550/arXiv.2408.12064}
  {10.48550/arXiv.2408.12064} (\bibinfo {year} {2024})\BibitemShut {NoStop}%
\bibitem [{\citenamefont {Sanders}\ \emph {et~al.}(2015)\citenamefont
  {Sanders}, \citenamefont {Wallman},\ and\ \citenamefont
  {Sanders}}]{Sanders2015}%
  \BibitemOpen
  \bibfield  {author} {\bibinfo {author} {\bibfnamefont {Y.~R.}\ \bibnamefont
  {Sanders}}, \bibinfo {author} {\bibfnamefont {J.~J.}\ \bibnamefont
  {Wallman}},\ and\ \bibinfo {author} {\bibfnamefont {B.~C.}\ \bibnamefont
  {Sanders}},\ }\bibfield  {title} {\bibinfo {title} {Bounding quantum gate
  error rate based on reported average fidelity},\ }\href
  {https://doi.org/10.1088/1367-2630/18/1/012002} {\bibfield  {journal}
  {\bibinfo  {journal} {New Journal of Physics}\ }\textbf {\bibinfo {volume}
  {18}},\ \bibinfo {pages} {012002} (\bibinfo {year} {2015})}\BibitemShut
  {NoStop}%
\bibitem [{\citenamefont {Nakajima}\ \emph {et~al.}(2020)\citenamefont
  {Nakajima}, \citenamefont {Noiri}, \citenamefont {Kawasaki}, \citenamefont
  {Yoneda}, \citenamefont {Stano}, \citenamefont {Amaha}, \citenamefont
  {Otsuka}, \citenamefont {Takeda}, \citenamefont {Delbecq}, \citenamefont
  {Allison} \emph {et~al.}}]{Nakajima2020}%
  \BibitemOpen
  \bibfield  {author} {\bibinfo {author} {\bibfnamefont {T.}~\bibnamefont
  {Nakajima}}, \bibinfo {author} {\bibfnamefont {A.}~\bibnamefont {Noiri}},
  \bibinfo {author} {\bibfnamefont {K.}~\bibnamefont {Kawasaki}}, \bibinfo
  {author} {\bibfnamefont {J.}~\bibnamefont {Yoneda}}, \bibinfo {author}
  {\bibfnamefont {P.}~\bibnamefont {Stano}}, \bibinfo {author} {\bibfnamefont
  {S.}~\bibnamefont {Amaha}}, \bibinfo {author} {\bibfnamefont
  {T.}~\bibnamefont {Otsuka}}, \bibinfo {author} {\bibfnamefont
  {K.}~\bibnamefont {Takeda}}, \bibinfo {author} {\bibfnamefont {M.~R.}\
  \bibnamefont {Delbecq}}, \bibinfo {author} {\bibfnamefont {G.}~\bibnamefont
  {Allison}}, \emph {et~al.},\ }\bibfield  {title} {\bibinfo {title} {Coherence
  of a driven electron spin qubit actively decoupled from quasistatic noise},\
  }\href {https://doi.org/https://doi.org/10.1103/PhysRevX.10.011060}
  {\bibfield  {journal} {\bibinfo  {journal} {Physical Review X}\ }\textbf
  {\bibinfo {volume} {10}},\ \bibinfo {pages} {011060} (\bibinfo {year}
  {2020})}\BibitemShut {NoStop}%
\bibitem [{\citenamefont {Liu}\ \emph {et~al.}(2024)\citenamefont {Liu},
  \citenamefont {Wang}, \citenamefont {Sheffer},\ and\ \citenamefont
  {Wang}}]{Liu2024}%
  \BibitemOpen
  \bibfield  {author} {\bibinfo {author} {\bibfnamefont {B.-J.}\ \bibnamefont
  {Liu}}, \bibinfo {author} {\bibfnamefont {Y.-Y.}\ \bibnamefont {Wang}},
  \bibinfo {author} {\bibfnamefont {T.}~\bibnamefont {Sheffer}},\ and\ \bibinfo
  {author} {\bibfnamefont {C.}~\bibnamefont {Wang}},\ }\bibfield  {title}
  {\bibinfo {title} {Observation of discrete charge states of a coherent
  two-level system in a superconducting qubit}\ }\href
  {https://doi.org/10.48550/arXiv.2401.12183} {10.48550/arXiv.2401.12183}
  (\bibinfo {year} {2024})\BibitemShut {NoStop}%
\bibitem [{\citenamefont {Ye}\ \emph {et~al.}(2024)\citenamefont {Ye},
  \citenamefont {Ellaboudy},\ and\ \citenamefont {Nichol}}]{Ye2024}%
  \BibitemOpen
  \bibfield  {author} {\bibinfo {author} {\bibfnamefont {F.}~\bibnamefont
  {Ye}}, \bibinfo {author} {\bibfnamefont {A.}~\bibnamefont {Ellaboudy}},\ and\
  \bibinfo {author} {\bibfnamefont {J.~M.}\ \bibnamefont {Nichol}},\ }\bibfield
   {title} {\bibinfo {title} {Stabilizing an individual charge fluctuator in a
  {Si/SiGe} quantum dot}\ }\href {https://doi.org/10.48550/arXiv.2407.05439}
  {10.48550/arXiv.2407.05439} (\bibinfo {year} {2024})\BibitemShut {NoStop}%
\bibitem [{\citenamefont {Paquelet~Wuetz}\ \emph {et~al.}(2023)\citenamefont
  {Paquelet~Wuetz}, \citenamefont {Degli~Esposti}, \citenamefont {Zwerver},
  \citenamefont {Amitonov}, \citenamefont {Botifoll}, \citenamefont {Arbiol},
  \citenamefont {Sammak}, \citenamefont {Vandersypen}, \citenamefont {Russ},\
  and\ \citenamefont {Scappucci}}]{paquelet2023reducing}%
  \BibitemOpen
  \bibfield  {author} {\bibinfo {author} {\bibfnamefont {B.}~\bibnamefont
  {Paquelet~Wuetz}}, \bibinfo {author} {\bibfnamefont {D.}~\bibnamefont
  {Degli~Esposti}}, \bibinfo {author} {\bibfnamefont {A.-M.~J.}\ \bibnamefont
  {Zwerver}}, \bibinfo {author} {\bibfnamefont {S.~V.}\ \bibnamefont
  {Amitonov}}, \bibinfo {author} {\bibfnamefont {M.}~\bibnamefont {Botifoll}},
  \bibinfo {author} {\bibfnamefont {J.}~\bibnamefont {Arbiol}}, \bibinfo
  {author} {\bibfnamefont {A.}~\bibnamefont {Sammak}}, \bibinfo {author}
  {\bibfnamefont {L.~M.}\ \bibnamefont {Vandersypen}}, \bibinfo {author}
  {\bibfnamefont {M.}~\bibnamefont {Russ}},\ and\ \bibinfo {author}
  {\bibfnamefont {G.}~\bibnamefont {Scappucci}},\ }\bibfield  {title} {\bibinfo
  {title} {Reducing charge noise in quantum dots by using thin silicon quantum
  wells},\ }\href {https://doi.org/https://doi.org/10.1038/s41467-023-36951-w}
  {\bibfield  {journal} {\bibinfo  {journal} {Nature communications}\ }\textbf
  {\bibinfo {volume} {14}},\ \bibinfo {pages} {1385} (\bibinfo {year}
  {2023})}\BibitemShut {NoStop}%
\bibitem [{\citenamefont {Rojas-Arias}\ \emph {et~al.}()\citenamefont
  {Rojas-Arias}, \citenamefont {Kojima}, \citenamefont {Takeda}, \citenamefont
  {Stano}, \citenamefont {Nakajima}, \citenamefont {Yoneda}, \citenamefont
  {Noiri}, \citenamefont {Kobayashi}, \citenamefont {Loss},\ and\ \citenamefont
  {Tarucha}}]{rojas2024}%
  \BibitemOpen
  \bibfield  {author} {\bibinfo {author} {\bibfnamefont {J.~S.}\ \bibnamefont
  {Rojas-Arias}}, \bibinfo {author} {\bibfnamefont {Y.}~\bibnamefont {Kojima}},
  \bibinfo {author} {\bibfnamefont {K.}~\bibnamefont {Takeda}}, \bibinfo
  {author} {\bibfnamefont {P.}~\bibnamefont {Stano}}, \bibinfo {author}
  {\bibfnamefont {T.}~\bibnamefont {Nakajima}}, \bibinfo {author}
  {\bibfnamefont {J.}~\bibnamefont {Yoneda}}, \bibinfo {author} {\bibfnamefont
  {A.}~\bibnamefont {Noiri}}, \bibinfo {author} {\bibfnamefont
  {T.}~\bibnamefont {Kobayashi}}, \bibinfo {author} {\bibfnamefont
  {D.}~\bibnamefont {Loss}},\ and\ \bibinfo {author} {\bibfnamefont
  {S.}~\bibnamefont {Tarucha}},\ }\bibfield  {title} {\bibinfo {title} {The
  origins of noise in the {Zeeman} splitting of spin qubits in natural-silicon
  devices}\ }\href {https://doi.org/10.48550/arXiv.2408.13707}
  {10.48550/arXiv.2408.13707}\BibitemShut {NoStop}%
\bibitem [{\citenamefont {Magesan}\ \emph {et~al.}(2012)\citenamefont
  {Magesan}, \citenamefont {Gambetta}, \citenamefont {Johnson}, \citenamefont
  {Ryan}, \citenamefont {Chow}, \citenamefont {Merkel}, \citenamefont
  {da~Silva}, \citenamefont {Keefe}, \citenamefont {Rothwell}, \citenamefont
  {Ohki}, \citenamefont {Ketchen},\ and\ \citenamefont
  {Steffen}}]{Magesan2012}%
  \BibitemOpen
  \bibfield  {author} {\bibinfo {author} {\bibfnamefont {E.}~\bibnamefont
  {Magesan}}, \bibinfo {author} {\bibfnamefont {J.~M.}\ \bibnamefont
  {Gambetta}}, \bibinfo {author} {\bibfnamefont {B.~R.}\ \bibnamefont
  {Johnson}}, \bibinfo {author} {\bibfnamefont {C.~A.}\ \bibnamefont {Ryan}},
  \bibinfo {author} {\bibfnamefont {J.~M.}\ \bibnamefont {Chow}}, \bibinfo
  {author} {\bibfnamefont {S.~T.}\ \bibnamefont {Merkel}}, \bibinfo {author}
  {\bibfnamefont {M.~P.}\ \bibnamefont {da~Silva}}, \bibinfo {author}
  {\bibfnamefont {G.~A.}\ \bibnamefont {Keefe}}, \bibinfo {author}
  {\bibfnamefont {M.~B.}\ \bibnamefont {Rothwell}}, \bibinfo {author}
  {\bibfnamefont {T.~A.}\ \bibnamefont {Ohki}}, \bibinfo {author}
  {\bibfnamefont {M.~B.}\ \bibnamefont {Ketchen}},\ and\ \bibinfo {author}
  {\bibfnamefont {M.}~\bibnamefont {Steffen}},\ }\bibfield  {title} {\bibinfo
  {title} {Efficient measurement of quantum gate error by interleaved
  randomized benchmarking},\ }\href
  {https://doi.org/10.1103/physrevlett.109.080505} {\bibfield  {journal}
  {\bibinfo  {journal} {Physical Review Letters}\ }\textbf {\bibinfo {volume}
  {109}},\ \bibinfo {pages} {080505} (\bibinfo {year} {2012})}\BibitemShut
  {NoStop}%
\bibitem [{\citenamefont {Huber}(1981)}]{huber_robust_1981}%
  \BibitemOpen
  \bibfield  {author} {\bibinfo {author} {\bibfnamefont {P.~J.}\ \bibnamefont
  {Huber}},\ }\href@noop {} {\emph {\bibinfo {title} {Robust statistics}}},\
  Wiley series in probability and mathematical statistics\ (\bibinfo
  {publisher} {J. Wiley \& sons},\ \bibinfo {address} {New York},\ \bibinfo
  {year} {1981})\BibitemShut {NoStop}%
\end{thebibliography}%


\begin{thebibliography}{4}%
\makeatletter
\providecommand \@ifxundefined [1]{%
 \@ifx{#1\undefined}
}%
\providecommand \@ifnum [1]{%
 \ifnum #1\expandafter \@firstoftwo
 \else \expandafter \@secondoftwo
 \fi
}%
\providecommand \@ifx [1]{%
 \ifx #1\expandafter \@firstoftwo
 \else \expandafter \@secondoftwo
 \fi
}%
\providecommand \natexlab [1]{#1}%
\providecommand \enquote  [1]{``#1''}%
\providecommand \bibnamefont  [1]{#1}%
\providecommand \bibfnamefont [1]{#1}%
\providecommand \citenamefont [1]{#1}%
\providecommand \href@noop [0]{\@secondoftwo}%
\providecommand \href [0]{\begingroup \@sanitize@url \@href}%
\providecommand \@href[1]{\@@startlink{#1}\@@href}%
\providecommand \@@href[1]{\endgroup#1\@@endlink}%
\providecommand \@sanitize@url [0]{\catcode `\\12\catcode `\$12\catcode
  `\&12\catcode `\#12\catcode `\^12\catcode `\_12\catcode `\%12\relax}%
\providecommand \@@startlink[1]{}%
\providecommand \@@endlink[0]{}%
\providecommand \url  [0]{\begingroup\@sanitize@url \@url }%
\providecommand \@url [1]{\endgroup\@href {#1}{\urlprefix }}%
\providecommand \urlprefix  [0]{URL }%
\providecommand \Eprint [0]{\href }%
\providecommand \doibase [0]{https://doi.org/}%
\providecommand \selectlanguage [0]{\@gobble}%
\providecommand \bibinfo  [0]{\@secondoftwo}%
\providecommand \bibfield  [0]{\@secondoftwo}%
\providecommand \translation [1]{[#1]}%
\providecommand \BibitemOpen [0]{}%
\providecommand \bibitemStop [0]{}%
\providecommand \bibitemNoStop [0]{.\EOS\space}%
\providecommand \EOS [0]{\spacefactor3000\relax}%
\providecommand \BibitemShut  [1]{\csname bibitem#1\endcsname}%
\let\auto@bib@innerbib\@empty
\bibitem [{\citenamefont {Rojas-Arias}\ \emph {et~al.}()\citenamefont
  {Rojas-Arias}, \citenamefont {Kojima}, \citenamefont {Takeda}, \citenamefont
  {Stano}, \citenamefont {Nakajima}, \citenamefont {Yoneda}, \citenamefont
  {Noiri}, \citenamefont {Kobayashi}, \citenamefont {Loss},\ and\ \citenamefont
  {Tarucha}}]{rojas2024}%
  \BibitemOpen
  \bibfield  {author} {\bibinfo {author} {\bibfnamefont {J.~S.}\ \bibnamefont
  {Rojas-Arias}}, \bibinfo {author} {\bibfnamefont {Y.}~\bibnamefont {Kojima}},
  \bibinfo {author} {\bibfnamefont {K.}~\bibnamefont {Takeda}}, \bibinfo
  {author} {\bibfnamefont {P.}~\bibnamefont {Stano}}, \bibinfo {author}
  {\bibfnamefont {T.}~\bibnamefont {Nakajima}}, \bibinfo {author}
  {\bibfnamefont {J.}~\bibnamefont {Yoneda}}, \bibinfo {author} {\bibfnamefont
  {A.}~\bibnamefont {Noiri}}, \bibinfo {author} {\bibfnamefont
  {T.}~\bibnamefont {Kobayashi}}, \bibinfo {author} {\bibfnamefont
  {D.}~\bibnamefont {Loss}},\ and\ \bibinfo {author} {\bibfnamefont
  {S.}~\bibnamefont {Tarucha}},\ }\bibfield  {title} {\bibinfo {title} {The
  origins of noise in the {Zeeman} splitting of spin qubits in natural-silicon
  devices}\ }\href {https://doi.org/10.48550/arXiv.2408.13707}
  {10.48550/arXiv.2408.13707}\BibitemShut {NoStop}%
\bibitem [{\citenamefont {Paquelet~Wuetz}\ \emph {et~al.}(2023)\citenamefont
  {Paquelet~Wuetz}, \citenamefont {Degli~Esposti}, \citenamefont {Zwerver},
  \citenamefont {Amitonov}, \citenamefont {Botifoll}, \citenamefont {Arbiol},
  \citenamefont {Sammak}, \citenamefont {Vandersypen}, \citenamefont {Russ},\
  and\ \citenamefont {Scappucci}}]{paquelet2023reducing}%
  \BibitemOpen
  \bibfield  {author} {\bibinfo {author} {\bibfnamefont {B.}~\bibnamefont
  {Paquelet~Wuetz}}, \bibinfo {author} {\bibfnamefont {D.}~\bibnamefont
  {Degli~Esposti}}, \bibinfo {author} {\bibfnamefont {A.-M.~J.}\ \bibnamefont
  {Zwerver}}, \bibinfo {author} {\bibfnamefont {S.~V.}\ \bibnamefont
  {Amitonov}}, \bibinfo {author} {\bibfnamefont {M.}~\bibnamefont {Botifoll}},
  \bibinfo {author} {\bibfnamefont {J.}~\bibnamefont {Arbiol}}, \bibinfo
  {author} {\bibfnamefont {A.}~\bibnamefont {Sammak}}, \bibinfo {author}
  {\bibfnamefont {L.~M.}\ \bibnamefont {Vandersypen}}, \bibinfo {author}
  {\bibfnamefont {M.}~\bibnamefont {Russ}},\ and\ \bibinfo {author}
  {\bibfnamefont {G.}~\bibnamefont {Scappucci}},\ }\bibfield  {title} {\bibinfo
  {title} {Reducing charge noise in quantum dots by using thin silicon quantum
  wells},\ }\href {https://doi.org/https://doi.org/10.1038/s41467-023-36951-w}
  {\bibfield  {journal} {\bibinfo  {journal} {Nature communications}\ }\textbf
  {\bibinfo {volume} {14}},\ \bibinfo {pages} {1385} (\bibinfo {year}
  {2023})}\BibitemShut {NoStop}%
\bibitem [{\citenamefont {Magesan}\ \emph {et~al.}(2012)\citenamefont
  {Magesan}, \citenamefont {Gambetta}, \citenamefont {Johnson}, \citenamefont
  {Ryan}, \citenamefont {Chow}, \citenamefont {Merkel}, \citenamefont
  {da~Silva}, \citenamefont {Keefe}, \citenamefont {Rothwell}, \citenamefont
  {Ohki}, \citenamefont {Ketchen},\ and\ \citenamefont
  {Steffen}}]{Magesan2012}%
  \BibitemOpen
  \bibfield  {author} {\bibinfo {author} {\bibfnamefont {E.}~\bibnamefont
  {Magesan}}, \bibinfo {author} {\bibfnamefont {J.~M.}\ \bibnamefont
  {Gambetta}}, \bibinfo {author} {\bibfnamefont {B.~R.}\ \bibnamefont
  {Johnson}}, \bibinfo {author} {\bibfnamefont {C.~A.}\ \bibnamefont {Ryan}},
  \bibinfo {author} {\bibfnamefont {J.~M.}\ \bibnamefont {Chow}}, \bibinfo
  {author} {\bibfnamefont {S.~T.}\ \bibnamefont {Merkel}}, \bibinfo {author}
  {\bibfnamefont {M.~P.}\ \bibnamefont {da~Silva}}, \bibinfo {author}
  {\bibfnamefont {G.~A.}\ \bibnamefont {Keefe}}, \bibinfo {author}
  {\bibfnamefont {M.~B.}\ \bibnamefont {Rothwell}}, \bibinfo {author}
  {\bibfnamefont {T.~A.}\ \bibnamefont {Ohki}}, \bibinfo {author}
  {\bibfnamefont {M.~B.}\ \bibnamefont {Ketchen}},\ and\ \bibinfo {author}
  {\bibfnamefont {M.}~\bibnamefont {Steffen}},\ }\bibfield  {title} {\bibinfo
  {title} {Efficient measurement of quantum gate error by interleaved
  randomized benchmarking},\ }\href
  {https://doi.org/10.1103/physrevlett.109.080505} {\bibfield  {journal}
  {\bibinfo  {journal} {Physical Review Letters}\ }\textbf {\bibinfo {volume}
  {109}},\ \bibinfo {pages} {080505} (\bibinfo {year} {2012})}\BibitemShut
  {NoStop}%
\bibitem [{\citenamefont {Huber}(1981)}]{huber_robust_1981}%
  \BibitemOpen
  \bibfield  {author} {\bibinfo {author} {\bibfnamefont {P.~J.}\ \bibnamefont
  {Huber}},\ }\href@noop {} {\emph {\bibinfo {title} {Robust statistics}}},\
  Wiley series in probability and mathematical statistics\ (\bibinfo
  {publisher} {J. Wiley \& sons},\ \bibinfo {address} {New York},\ \bibinfo
  {year} {1981})\BibitemShut {NoStop}%
\end{thebibliography}%
\end{document}


\beginsupplement
	
	
	
	\title{Supplemental Material for ``Efficient Qubit Calibration \\ by Binary-Search Hamiltonian Tracking''}
	
	\author{Fabrizio~Berritta}
        \affiliation{\RLEaffil}	
        \affiliation{\QDevaffil}	
	\author{Jacob~Benestad}
	   \affiliation{\NTNUaffil}
    \author{Lukas~Pahl}
        \affiliation{\RLEaffil}
        \affiliation{\EECSaffil}
    \author{Melvin Mathews}
        \affiliation{\RLEaffil}
        \affiliation{Department of Information Technology and Electrical Engineering, ETH Z\"urich, 8093 Z\"urich, Switzerland}
	\author{Jan~A.~Krzywda}
	   \affiliation{$\langle a Q a^L \rangle$ Applied Quantum Algorithms --- Lorentz Institute for Theoretical Physics \& Leiden Institute of Advanced Computer Science, Universiteit Leiden, The Netherlands}
    \author{R\'eouven~Assouly}
        \affiliation{\RLEaffil}
    \author{Youngkyu~Sung}
        \affiliation{\RLEaffil}
        \affiliation{\EECSaffil}
    \author{David~K.~Kim}
        \affiliation{\LLaffil}
    \author{Bethany~M.~Niedzielski}
        \affiliation{\LLaffil}
    \author{Kyle~Serniak}
        \affiliation{\RLEaffil}
        \affiliation{\LLaffil}
    \author{Mollie~E.~Schwartz}
        \affiliation{\LLaffil}
    \author{Jonilyn~L.~Yoder}
        \affiliation{\LLaffil}
	\author{Anasua~Chatterjee}
	   \affiliation{\QDevaffil}	
        \affiliation{
        QuTech and Kavli Institute of Nanoscience, Delft University of Technology, Delft, The Netherlands}	
    \author{Jeffrey~A.~Grover}
        \affiliation{\RLEaffil}
    \author{Jeroen~Danon}
        \affiliation{\NTNUaffil}
    \author{William~D.~Oliver}
        \affiliation{\RLEaffil}
        \affiliation{\EECSaffil}
        \affiliation{\Physaffil}
    \author{Ferdinand~Kuemmeth}
        \email{ferdinand.kuemmeth@ur.de}
	   \affiliation{\QDevaffil}	
        \affiliation{Institute of Experimental and Applied Physics, University of Regensburg, 93040 Regensburg, Germany}
	   \affiliation{QDevil, Quantum Machines, 2750 Ballerup, Denmark}

	
	\date{August 26, 2025}
	\maketitle
	\tableofcontents

\section{Experimental setup}
\begin{figure*}[h]
	\centering
	\includegraphics[height=0.9\textheight,keepaspectratio]{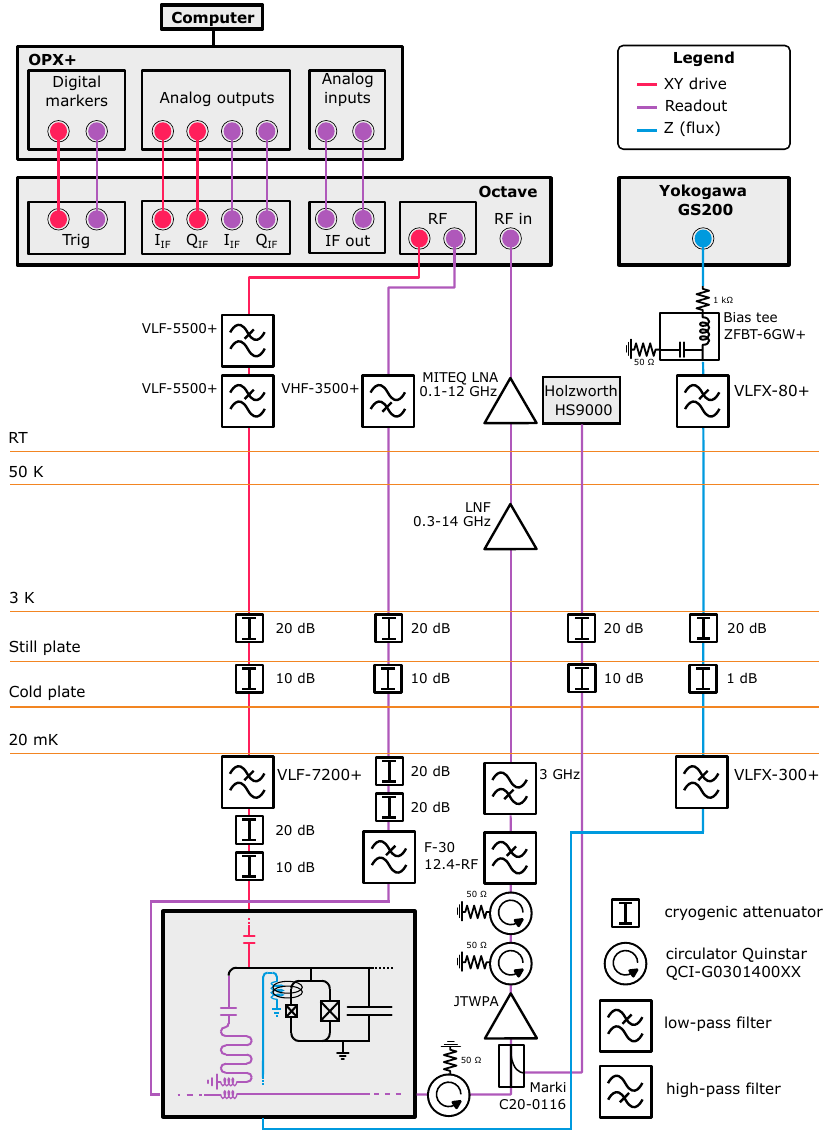}
	\caption[Experimental setup BF2]{\textbf{Experimental setup BF2.} The cryostat is a Bluefors XLD-600 dilution refrigerator with a base temperature lower than 30 mK. A Quantum Machines OPX+ and Octave are used for the XY drive pulses and readout. The setup includes a Yokogawa GS200 for DC flux biasing.}
 \label{fig:FigS1_FBS}
\end{figure*}
The measurements are performed in a Bluefors XLD-600 dilution refrigerator with a base temperature of $\SI{20}{\milli\kelvin}$ and the setup is sketched in Fig.~\ref{fig:FigS1_FBS}. The Quantum Machines OPX+ and Octave are used for the XY control of the qubit and readout signal, and both microwave pulses are generated by single-sideband modulation with suppressed carrier. For the qubit coherence experiment in the main text, each drive pulse is 28-ns long, with a cosine rise and fall envelope of $\SI{5}{\nano\second}$ each. For the randomized benchmarking and gate set tomography experiments, the pulses are 20-ns long and calibrated by DRAG.

The OPX+ includes real-time classical processing with fast analog feedback programmed in QUA software. The OPX+ and Octave are frequency-locked by a Quantum Machines OPT (not shown). 
The readout resonator linewidth $\kappa_\text{r}/(2\pi)\approx \SI{0.35}{\mega\hertz}$, and the microwave readout tone, approximately $\SI{7.08}{\giga\hertz}$, is filtered and attenuated at room temperature by passive components. The readout tone is attenuated in the cryostat to remove excess thermal photons from higher-temperature stages, and low-pass filtered at the mixing chamber. The device is shielded magnetically with a superconducting can. The transmitted signal from the feedline goes through a circulator placed after the sample to remove noise of the Josephson traveling wave parametric amplifier (JTWPA) and dump it in a $\SI{50}{\ohm}$ terminator. We preamplify the readout signal at base temperature using a Holzworth HS9000 Synthesizer. 

To prevent noise from higher-temperature stages from reaching the JTWPA and the sample, two microwave isolators are placed after the JTWPA. A high-electron mobility transistor amplifier thermally anchored at the  $\SI{3}{\kelvin}$ stage further amplifies the readout signal. At room temperature, the readout line is again amplified. A Yokogawa GS200 provides the DC bias flux through a bias-tee terminated by $\SI{50}{\ohm}$ from the AC side, as we perform virtual Z gates in this work.
\begin{table}[H]
    \centering
    \begin{tabular}{l l c}
        \toprule
        \toprule
        \multicolumn{3}{c}{\textbf{Device parameters}} \\
        \midrule
        Transition frequency & $\omega_\text{q}/(2\pi)$ ($\SI{}{\giga\hertz}$) & 3.78  \\
        \midrule
        Anharmonicity & $\eta/(2\pi)$ ($\SI{}{\mega\hertz}$)  & -248 \\
        \midrule
        Relaxation time & $T_1$ ($\SI{}{\micro\second}$) & 80 \\
        \midrule
        Ramsey decay time & $T_2^*\textsuperscript{a}$ ($\SI{}{\micro\second}$) & 3.3 \\
        \midrule
        Spin-echo decay time & $T_{2\text{E}}$ ($\SI{}{\micro\second}$) & 12 \\
        \midrule
        Ramsey decay time ($\Phi_{\rm ext} = 0$) & $T_2^*\textsuperscript{b}$ ($\SI{}{\micro\second}$) & 42 \\
        \midrule
        Readout resonator frequency & $\omega_\text{r}/(2\pi)$ ($\SI{}{\giga\hertz}$) & 7.08  \\
        \midrule
        Readout resonator linewidth & $\kappa_\text{r}/(2\pi)$ ($\SI{}{\mega\hertz}$)  & 0.35 \\
        \bottomrule
        \bottomrule
    \end{tabular}
    
    \caption{Device parameters at $\Phi_{\text{ext}}=\Phi_0/4$ if not specified otherwise.\\
    \protect\textsuperscript{a} Measured over six hours. \\
    \protect\textsuperscript{b} Measured over one minute.}
    \label{tab:noise}
\end{table}

\clearpage

\section{The frequency binary search algorithm}

Here we summarize the FBS algorithm described in the main text, together with a concise pseudocode format as seen in Algorithm~\ref{alg:fbs}. The inputs to the algorithm are $\mu_0$ and $\sigma_0$ to parameterize the initial Gaussian prior (i.e., roughly the search area for the frequency $\varepsilon$ to be estimated), in addition to the number of update steps $N$ and SPAM error and dephasing coefficients $\alpha$, $\beta$ and $T$. Each update step starts by calculating the qubit precession time $\tau$ to balance that the oscillation period of the likelihood is closest possible to the width (specifically $2\pi\sigma$) of the prior distribution while also not reaching so long times that more information is lost through dephasing than necessary (formally, one calculates the $\tau$ that minimizes the expected posterior variance). Based on $\tau$, the frequency detuning $\Updelta f$ is then calculated so that the inflection point of the likelihood function is located in the middle of the prior. This combination of $\tau$ and $\Updelta f$ divides the prior in a left and right branch based on the outcome of a Ramsey measurement $m\in\{-1,1\}$. The parameters $\mu$ and $\sigma$ for the Gaussian used to approximate the true posterior distribution are then calculated based on a method of moments fit where these parameters are taken to be the mean and standard deviation of the true posterior. This update sequence is repeated, now with $\mu$ and $\sigma$ parameterizing the new prior, until it has been performed a total of $N$ times.

\begin{algorithm}[H]
\caption{Frequency binary search algorithm}\label{alg:fbs}
\begin{algorithmic}[1]
\Procedure{FrequencyBinarySearch}{$\mu_0, \sigma_0, N,\alpha,\beta,T$}

\State $\mu \gets \mu_0$
\State $\sigma \gets \sigma_0$
\State $n \gets 0$
\For{$n<N$}
\State $\tau \gets \frac{\sqrt{16\pi^2\sigma^2+1/T^2}-1/T}{8\pi^2\sigma^2}$
\State $\Updelta f \gets \frac{1}{4\tau} - \mu$
\State $m \gets \Call{Ramsey}{\tau, \Updelta f}$ \Comment{$m\in\{-1,1\}$ is the experiment outcome}
\State $\mu \gets \mu-\frac{2\pi m\beta\sigma^2\tau e^{-\tau/T-2\pi^2\sigma^2\tau^2}}{1+m \alpha}$
\State $\sigma \gets \sqrt{\sigma^2-\frac{4\pi^2\beta^2\sigma^4\tau^2e^{-2\tau/T-4\pi^2\sigma^2\tau^2}}{(1+m\alpha)^2}}$
\State $n \gets n+1$
\EndFor

\State \textbf{return} $\mu$, $\sigma$
\EndProcedure
\end{algorithmic}
\end{algorithm}

\clearpage

\clearpage
\section{Qubit stabilization over six hours}
\begin{figure}[h]
	\centering
	\includegraphics{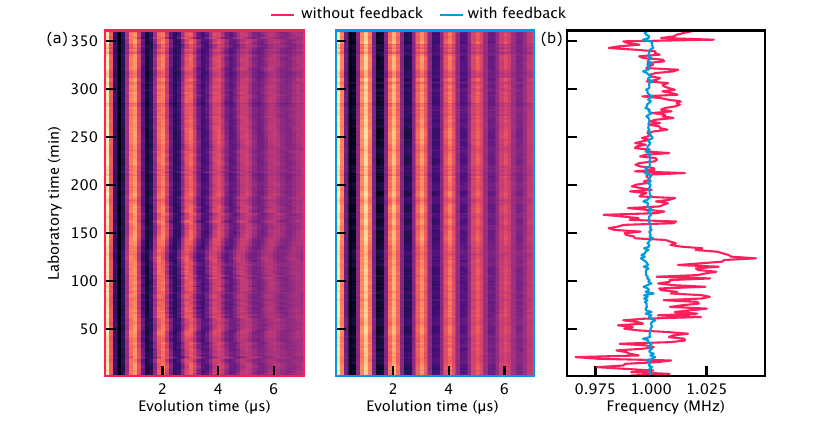}
	\caption[Qubit frequency stabilization over 6 hours] {\textbf{Qubit frequency stabilization over 6 hours.}
		\textbf{(a)} The Ramsey experiment is repeated for 6 hours without (left) and with (right) feedback by the frequency binary search. With the feedback, the qubit frequency is locked during the whole duration of the experiment.
		\textbf{(b)} Time dependence of the measured frequency detuning extracted from Ramsey measurements in (a). The red trace is taken with a fixed drive frequency $f_\text{d}$, and the blue trace is taken with a feedback-controlled $f_\text{d}$.
	}
	\label{fig:FigS2_FBS} 
\end{figure}

We perform the protocol explained in Fig.~2 of the main text for hours as shown for the interleaved measurements in Fig.~\ref{fig:FigS2_FBS}(a), except this time the controller uses $N=6$ single-shot measurements for the FBS estimation instead of $N=8$. In the left (right) panel the feedback is turned off (on) for over 6 hours. Figure~\ref{fig:FigS2_FBS}(b) shows the values of $\Updelta f$ obtained from Ramsey measurements with feedback off (red curve) and on (blue curve). The feedback significantly reduces the fluctuation of $\Updelta f$. By averaging across laboratory time panels (a) and (b), $T_2^*$ improves from $\SI{3.32(0.10)}{\micro\second}$ without feedback to $\SI{4.81(0.13)}{\micro\second}$ with feedback, which results in an average 44\% improvement, similar to Fig.~2 of the main text.

\subsection{Power spectral density with and without feedback}
We use the 6-hour-long Ramsey measurements to extract the power spectral density (PSD) of the qubit frequency fluctuations with and without the feedback. We combine four different methods to extract the PSD across seven orders of magnitude (the result is shown in Fig.~\ref{Fig:figSpsd_psd}):
\begin{figure}[t]
    \centering
    \includegraphics{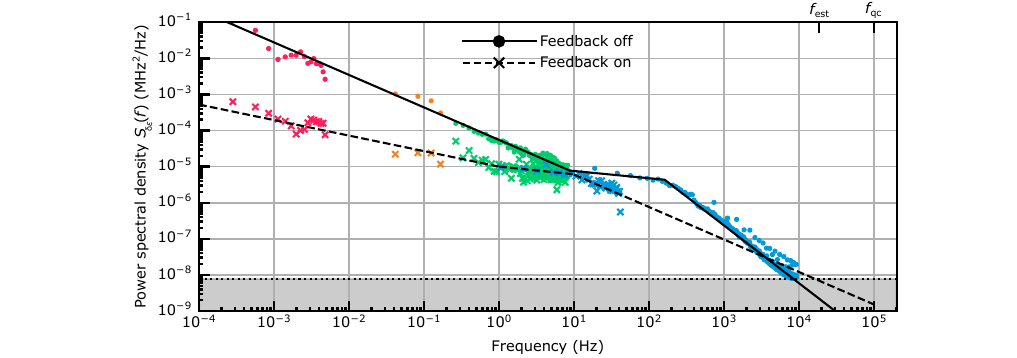}
    \caption{\textbf{Power Spectral Density of frequency fluctuations with and without feedback.} We use four different methods to process 6 hours of Ramsey data and estimate the PSD across eight (five) orders of magnitude for data without (with) feedback. We perform a piecewise fit to a power-law spectrum, i.e., $S_f(\delta \epsilon) = A (f_1/f)^\alpha$, yielding three different regions for the data without feedback (drawn with solid black lines). For the data with feedback we use two regions (dashed black lines). The horizontal dotted line, denotes the white-noise floor corresponding to the experimentally measured $T_2 = 12\,\SI{}{\micro\second}$. We plot a one-sided PSD. The extracted parameters are given in Tab.~\ref{tab:psd_params}.}
    \label{Fig:figSpsd_psd}
\end{figure}
\begin{itemize}
\item
At low frequencies (red points), we use the frequency time-series from Fig.~\ref{fig:FigS2_FBS} and compute the PSD using Welch's method. In this approach, each data point is separated by an approximately constant delay of $104.37\,\text{s}$, which consists of a measurement period ($30.19\,\text{s}$) and data transfer ($74.18\,\text{s}$).

\item
At intermediate frequencies (orange points), we repeat this procedure inside each of 207 runs, using an averaging window of 1,000 repetitions, which effectively corresponds to $3.02\,\text{s}$ window duration. Empirically, we found that smaller averaging windows lead to artificial noise due to errors in fitting the frequency to the averaged oscillating signal.

\item
At high frequencies (green points), we adapt the method of Ref.~\cite{rojas2024}, and relate the spectral density of the frequency fluctuations to the correlation function of the initial state population:
\[
P_0(t;\tau) = \frac{1}{2} + \frac{1}{2}\bigg(\alpha + \beta e^{-\tau/T} \cos(2\pi [\epsilon_0+\delta\epsilon(t)]\tau)\bigg) \approx A_0(\tau) + B_1(\tau) \delta\epsilon(t) \tau +B_2(\tau)\delta\epsilon(t)^2 \tau^2,
\]
where
\begin{align*}
A_0(\tau) = {} & {}   \frac{1}{2} + \frac{\alpha}{2} + \frac{\beta e^{-\tau/T}}{2} \cos(2\pi\epsilon_0\tau),\\
B_1(\tau) = {} & {} -\pi \beta e^{-\tau/T} \sin(2\pi\epsilon_0\tau),\\
B_2(\tau) = {} & {}  -\pi^2 \beta e^{-\tau/T} \cos(2\pi\epsilon_0\tau),
\end{align*}
with $t$ denoting laboratory time, in contrast to $\tau$, which stands for a probing time. In this model, $\delta \epsilon(t)$ can be seen as the coarse-grained fluctuations in the qubit frequency, which, for considered probing times of $\tau < 7\,\SI{}{\micro\second}$ and expected noise amplitudes of $\sigma_{\delta \epsilon} \sim 0.01\,\text{MHz}$, results in relatively small random phases $\delta \phi = 2\pi\delta \epsilon(t) \tau \ll 1$, which validates the expansion. The correlation function of $P_0$ is now related to the correlation function of the frequency fluctuations $R_{\delta\epsilon}(t)= \langle \delta \epsilon(t) \delta \epsilon(0) \rangle$ as
\[
C_P(t;\tau) = \langle P_0(t) P_0(0) \rangle - \langle P_0(0) \rangle^2  \approx B_1^2 R_{\delta\epsilon}(t)\tau^2 + B_2^2 R_{\delta\epsilon}(t)^2 \tau^4.
\]
The above relation holds for any $\tau$; however, it is convenient to use a $\tau$ that (i) makes the $B_2 \propto \cos(2\pi \epsilon_0 \tau)$ term small, (ii) is large enough to provide a good signal-to-noise ratio, but (iii) is small enough not to be affected by relaxation. By optimizing the signal-to-noise ratio, we pick $\tau_0 = \{ 3.721, 3.864, 4.149 , 4.292\} \, \SI{}{\micro\second}$ probing times, which confirms the observation that the optimal $\tau_0 \approx T_2^*$ \cite{rojas2024}. At this point, we use the Wiener-Khinchin theorem to find the spectral density as the Fourier transform of the extracted correlation function, i.e.,
\begin{equation}
    S(f) = \int dt\, e^{-i2\pi f t} R_{\epsilon}(t) \approx \int dt \frac{\hat{C}_P(t;\tau_0)}{B_1^2 \tau_0^2} e^{-i2\pi f t},
\end{equation}
where $\hat{C}_P(t;\tau)$ is the estimated correlation function of the population, averaged over the 207 runs and waiting times $\tau_0$. In a single repetition, the correlation function can be empirically computed as
\begin{equation}
    \hat{C}_P(k\Delta t;\tau_0) = \frac{1}{N} \sum_{n=1}^N x_{\tau_0}[n] x_{\tau_0}[n+k] - \frac{1}{N^2} \bigg(\sum_{n=1}^N x_{\tau_0}[n]\bigg)^2
\end{equation}
where $x_{\tau_0}[n]  = \pm 1$ is the $n$-th measurement of the population at time $\tau_0$ and $\Delta t \approx 3.5$ ms is the time between two measurements. 

\item
Finally, at the highest frequencies (blue points), we rely on the time series of frequency fluctuations extracted from Bayesian estimation. For the data without feedback, we directly use the frequencies estimated in the FBS scheme, allowing us to extract the PSD up to the inverse of the estimation cycle $f_\text{max} \approx 10^4\,$Hz. For the noise with feedback, we perform offline Bayesian estimation using 4 repetitions, i.e., 200 probing times, which in total takes $\Delta t = 4 \times 3.5\,$ ms$\,=  14$ ms and sets $f_\text{max} \approx 60$~Hz. For the estimation, we use a Gaussian prior centered around the previous estimate with an initial $\sigma_0 = 2\,\text{kHz}$. Larger $\sigma_0$ resulted in an artificially flat spectrum due to large uncertainty in estimation, while smaller $\sigma_0$ introduced artificial low-pass filtering. For both the feedback and no-feedback data, we converted the frequency fluctuations time trace to a PSD using Welch's method.
\end{itemize}

Together, the extracted values of the PSDs are plotted in Fig.~\ref{Fig:figSpsd_psd} as a function of frequency. We use dots for no-feedback data and crosses for data with feedback, and use different colors for the four methods described above. Both PSDs are piecewise fitted with a power law $S(f) = A (f_1/f)^\alpha$, where $A$ and $\alpha$ are the fitting parameters and $f_1 = 1\,$Hz. The extracted values of $A$ and $\alpha$ are shown in Tab.~\ref{tab:psd_params}.

\begin{figure}[htb!]
    \centering
    \includegraphics[width=0.7\linewidth]{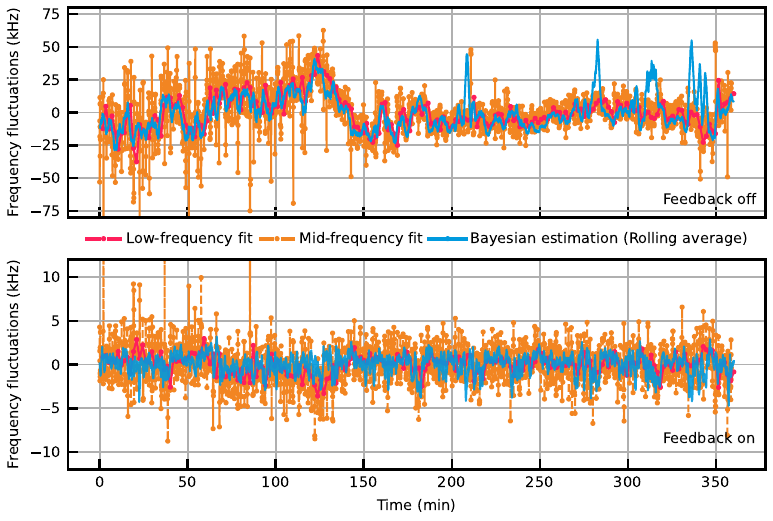}
    \caption{\textbf{Frequency fluctuations extracted from 6-hours Ramsey}. We cross-validate methods by plotting the frequency fluctuations as a function of laboratory time. For the Bayesian estimation data we apply a rolling average with $N = 10,000$ for the data without feedback and $N=1,000$ for the data with feedback. Furthermore we downsample the averaged Bayesian signal by a factor of $M=100$.}
    \label{Fig:figSpsd_time_ser} 
\end{figure}

To cross-check different methods of estimating power spectral density, we compare in Fig.~\ref{Fig:figSpsd_time_ser} time traces of frequency fluctuations obtained in the low- and mid-frequency methods against Bayesian estimation data with a moving average, obtaining relatively good agreement.

\subsection{Effect of the feedback on the non-Markovian dephasing}

Based on the extracted power spectral density, we can estimate the quasistatic dephasing time, using the relation \cite{paquelet2023reducing}
\begin{equation}\label{eq:t2star}
   T_2^* = \frac{1}{\sqrt{2}\pi \sigma_f}, \quad \sigma_f^2 = \int_{f_{0}}^{f_{\rm qc}} \text{d}f\, S(f),
\end{equation}
where $S(f)$ is the one-sided spectrum, i.e., $S(f) = S_2(f) + S_2(-f)$ in terms of the two-sided spectrum $S_2(f)$.
The lower bound $f_0$ is set by the duration of the experiment used to determine $T_2^*$.
For the upper bound $f_{\rm qc}$ we assumed that the PSD is dominated by frequencies below the inverse of the qubit cycle time $\approx 10\,\SI{}{\micro\second}$, yielding $f_{\rm qc} = 10^5\,\text{Hz}$ as upper cut-off frequency. 

For each part of the PSD that fits a power law $S(f) = A (f_1/f)^{\alpha}$ between frequencies $f_{\rm min}$ and $f_{\rm max}$, we can thus estimate its contribution to the noise amplitude as
\begin{equation}
    \int_{f_{\rm min}}^{f_{\rm max}} \text{d}f\, S(f) = \begin{cases}
     \frac{A f_1}{1-\alpha} (f_{\rm min}/f_1)^{1-\alpha} \left( \kappa^{1-\alpha} - 1 \right) & \text{for } \alpha \neq 1, \\
    A f_1 \ln(\kappa) & \text{for } \alpha = 1,
    \end{cases}\label{eq:intS}
\end{equation}
where $\kappa = f_{\rm max}/f_{\rm min}$.
The total $\sigma_f^2$ in Eq.~(\ref{eq:t2star}) can then be expressed as a sum of the contributions from different regions of the PSD using the piecewise fits shown in Fig.~\ref{Fig:figSpsd_psd}. We present the results for each piecewise region of the PSD in Tab.~\ref{tab:psd_params} below.

\setlength{\tabcolsep}{10pt} 
\begin{table}[htbp]
\centering
\caption{Integrated PSD ($\sigma_f^2$) and various $T_2$ times with feedback status and fit parameters. 
The $\sigma_f^2$ values are calculated using $\sigma^2 = \int_{f_{\text{min}}}^{f_{\text{max}}} A (f_1/f)^{\alpha} \text{d}f$.
The $T_2^*$ sums up the total contribution $1/(\sqrt{2}\pi\sqrt{\sum \sigma_i^2})$ from the different regions.
$T_{\rm E} = 12 \,\SI{}{\micro\second}$ is the measured Hahn echo time, used for estimating $\hat{T}_2$ in both groups, according to relation $1/\hat{T}_2 = 1/T_2^* + 1/T_{\rm E}$.} 
\label{tab:psd_params} 
\begin{tabular}{@{}lccccccccc@{}} 
\toprule
Feedback & $f_{\text{min}}$ & $f_{\text{max}}$ & $A$                   & $\alpha$ & $\sigma_f^2$           & $T_2^*$   & $T_{\rm E}$      & $\hat{T}_2$ & $T_2$      \\
         & (Hz)             & (Hz)             & (MHz$^2$/Hz)          &          & (MHz$^2$)              & ($\SI{}{\micro\second}$)  & ($\SI{}{\micro\second}$)   & ($\SI{}{\micro\second}$)    & ($\SI{}{\micro\second}$)   \\
\midrule
\multirow{3}{*}{\centering off} & $1.0 \times 10^{-4}$ & $8.9 \times 10^0$ & $0.55 \times 10^{-4}$ & $0.90$   & $4.664 \times 10^{-4}$ & \multirow{3}{*}{\centering $4.55$} & \multirow{3}{*}{\centering $12$} & \multirow{3}{*}{\centering $3.30$} & \multirow{3}{*}{\centering $3.4$} \\
         & $8.9 \times 10^0$  & $1.6 \times 10^2$ & $0.12 \times 10^{-4}$ & $0.20$   & $8.364 \times 10^{-4}$ &                               &                    &                    &                    \\
         & $1.6 \times 10^2$  & $1.0 \times 10^5$ & $1.48 \times 10^{-2}$ & $1.60$   & $1.147 \times 10^{-3}$ &                               &                    &                    &                    \\
\midrule
\multirow{3}{*}{\centering on}  & $1.0 \times 10^{-4}$ & $1.0 \times 10^0$ & $0.10 \times 10^{-4}$ & $0.43$   & $1.745 \times 10^{-5}$ & \multirow{3}{*}{\centering $7.13$} & \multirow{3}{*}{\centering $12$} & \multirow{3}{*}{\centering $4.47$} & \multirow{3}{*}{\centering $4.8$} \\
         & $1.0 \times 10^0$  & $1.0 \times 10^1$ & $0.10 \times 10^{-4}$ & $0.21$   & $6.539 \times 10^{-5}$ &                               &                    &                    &                    \\
         & $1.0 \times 10^1$  & $1.0 \times 10^5$ & $0.48 \times 10^{-4}$ & $0.90$   & $9.134 \times 10^{-4}$ &                               &                    &                    &                    \\
\bottomrule
\end{tabular}
\end{table}

We start from the low-frequency region $f < 10\,$Hz, in which feedback provided more than a factor of two reduction in noise root mean square, i.e., from $\sigma_\text{off} = 21.6$ kHz to $\sigma_\text{on} =9.1$ kHz. However, for the data with the feedback, finite exponent $\alpha_\text{on} > 0$ suggests that filtering properties are worse for low-frequency noise. One explanation is that a narrow prior in the FBS protocol leads to underestimating large deviations from the average frequency, which are statistically associated with low-frequency noise.

The extracted PSD suggests that decoherence is more likely to be dominated by higher frequencies $f> 10\,$Hz. In this regime the improvement is weaker, of the order of $40\%$, reflected in a reduction from $\sigma_{\text{off}} \approx 45$ kHz to $\sigma_\text{on} \approx31$~kHz. We highlight however that in this regime the extracted PSD relies on sparse data, which includes extrapolation of an initial trend extracted from Bayesian data (feedback on). Also for feedback-off data, the kink in the spectral density might represent a change in the dominating physical mechanism responsible for the fluctuations, but could also reflect too narrow prior distribution resulting in an apparent low-pass filter. 

In total, the estimated quasistatic dephasing times of $T_{2,\text{off}}^* = 4.55\,\SI{}{\micro\second}$ and $T_{2,\text{on}} = 7.13 \,\SI{}{\micro\second}$ are slightly longer than experimentally measured $T_{2,\text{off}} = 3.4\,\SI{}{\micro\second}$ and $T_{2,\text{on}} = 4.8\,\SI{}{\micro\second}$. As a consequence, the predicted feedback-related improvement of $56\%$ slightly exceeds the reported $44\%$. Most likely, this is due to dephasing caused by high-frequency noise, with a characteristic frequency above $f_{\text{qc}} = 10^5\,\text{Hz}$. Such an explanation is consistent with the larger overestimation of the data with feedback, in case of which a relatively larger portion of dephasing can be attributed to unfiltered high-frequency noise.

To further verify this hypothesis, we use a heuristic formula for estimated total dephasing time $\hat T_2^*$, i.e.
\begin{equation}
    \frac{1}{\hat T_2} = \frac{1}{T_2^*} + \frac{1}{T_{\rm E}}
\end{equation}
where the spin-echo time $T_{\rm E}$ is approximately equal to the dephasing time due to high-frequency noise. Substituting the experimentally measured $T_{\rm E} = 12\,\SI{}{\micro\second}$, the estimated $\hat T_2$ moves closer to experimental values, i.e., $\hat{T}_{2,\text{off}} \approx 3.30\,\SI{}{\micro\second}$ and $\hat{T}_{2,\text{on}} \approx 4.47\,\SI{}{\micro\second}$, both within $10\%$ error. Similarly, the predicted, feedback-related improvement of $\approx 35\%$ is consistent with the experiment. Its slightly smaller value is likely related to larger underestimation of dephasing time for the data with the feedback, caused by extrapolation of the PSD in the high frequency regime. 

\clearpage
\section{Interleaved randomized benchmarking}
\begin{figure}[h]
	\centering
	\includegraphics{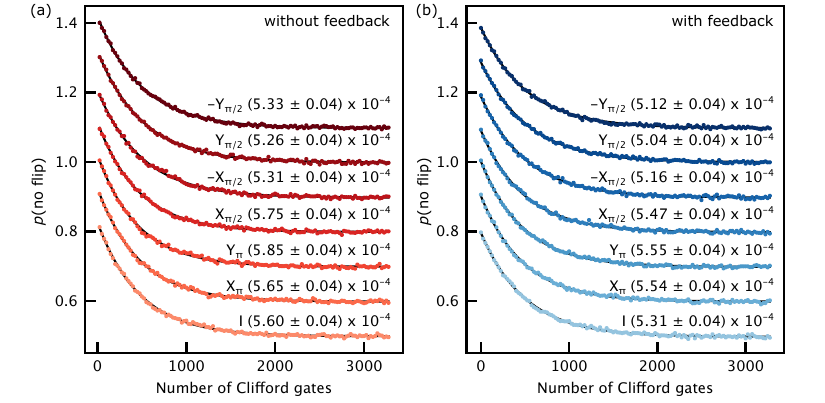}
	\caption[Interleaved randomized benchmarking with and without frequency binary search.] {\textbf{Interleaved randomized benchmarking with and without frequency binary search.}
		\textbf{(a)} Sequence infidelities for interleaved randomized benchmarking without frequency estimation by FBS. Traces are offset by 0.1 for clarity. Interleaved sequences are annotated with corresponding single-qubit gates and extracted natural gate infidelities.
		\textbf{(b)} Interleaved with (a) using frequency estimation by FBS. For all gates, the infidelity is slightly lower with feedback by a few $10^{-5}$.
	}
	\label{fig:FigS3_FBS} 
\end{figure}

We perform interleaved randomized benchmarking (RB)~\cite{Magesan2012} with and without feedback. The protocol consists of interleaving the gate $\mathcal{C}$ of interest with random gates from the Clifford sequence. A final gate is performed to make the total sequence equal to the identity operation. In Fig.~\ref{fig:FigS3_FBS} we plot the fraction of not-flipped state $p(\text{no flip)}$ as a function of the number of Clifford gates in a random sequence, terminated by a Clifford gate that would in principle bring the qubit back to the initial state. For each gate, 30 random sequences are generated and they are averaged 1,000 times. We find a X$_{\pi/2}$-pulse gate infidelity improvement from $(5.75\pm0.04)\times 10^{-4}$ to $(5.47\pm0.04)\times 10^{-4}$, and for the Y$_{\pi/2}$-pulse from $(5.26\pm0.04)\times 10^{-4}$ to $(5.04\pm0.04)\times 10^{-4}$. The interleaved RB infidelities are in the same order of magnitude as the standard RB shown in Fig.~3 of the main text. For each gate, feedback shows a similar improvement.

\clearpage
\section{On the impact of state and preparation measurement errors and dephasing}
\begin{figure}[h]
    \centering
    \includegraphics{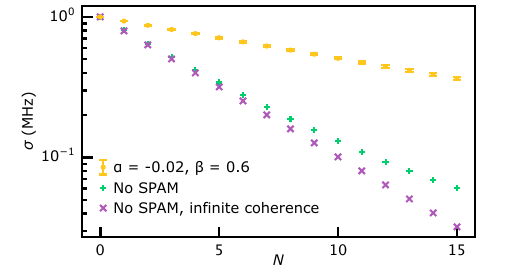}
    \caption{
    \textbf{Impact of SPAM errors and dephasing.} As the algorithm becomes aware of errors it becomes more conservative. As a reference we plot the posterior standard deviation $\sigma$ without SPAM errors (ie. $\alpha=0$ and $\beta=0$) both for $T\rightarrow +\infty$ (purple) and $T=\SI{10}{\micro\second}$ (green). For the case when $\alpha \neq 0$ we plot the average $\sigma$ over several simulations (yellow) with error bars (extending two standard deviations of $\sigma$ to each side), as it now becomes measurement dependent.}
    \label{fig:A2}
\end{figure}

In Fig.~\ref{fig:A2} we plot the scaling of the FBS uncertainty $\sigma$ as a function of the number of measurements $N$ for the case (i) no state and preparation measurement errors (SPAM), no decoherence, (ii) no SPAM, with decoherence $T=\SI{10}{\micro\second}$, and (iii) including SPAM and decoherence. By including SPAM of the type $\beta \neq 0$ and dephasing $T$ in the statistical model for how measurements are generated, the algorithm becomes more conservative in how much the posterior distribution variance is reduced as the visibility is decreased and the algorithm now has less trust in measurement outcomes. Having SPAM errors of the type $\alpha \neq 0$ reflects a bias towards one of the states being generated more than the other, and by making the algorithm aware of this bias it attempts to compensate by increasing the weight of the minority outcomes while reducing the weight of the majority outcomes.

The linlog plot in Fig.~\ref{fig:A2} of $\sigma$ versus $N$ demonstrates the scaling of the uncertainty in the algorithm. An exponential scaling of the posterior variance persists for as long as $\tau \ll T$, as the expected posterior variance in this case can be approximated as
\begin{equation}
\text{E}[\sigma^2] = \sigma^2 - \frac{4\pi^2\beta^2\tau^2\sigma^4}{1-\alpha^2} e^{-2\tau(1/T+2\pi^2\sigma^2\tau)} \approx \sigma^2 - \frac{4\pi^2\beta^2\tau^2\sigma^4}{1-\alpha^2} e^{-4\pi^2\sigma^2\tau^2},
\end{equation}
where the optimal time $\tau$ (presented in the main text) is now approximately
\begin{equation}
\tau \approx \frac{1}{2\pi\sigma},
\end{equation}
so that the variance scales as a constant factor
\begin{equation}
\text{E}[\sigma^2]  \approx \left(1 - \frac{\beta^2 e^{-1}}{1-\alpha^2} \right) \sigma^2 = \xi \sigma^2.
\end{equation}
As a function of $n$, this then gives the scaling
\begin{equation}
\sigma_n^2 \sim e^{n \ln{\xi}},
\end{equation}
where $\ln{\xi}<0$. A larger $\alpha$ and $\beta$ will have the effect of giving a slower exponential scaling by bringing $\xi$ closer to one, while $T$ serves as a sort of ``cutoff'' parameter that suppresses the exponential scaling when eventually $\tau$ becomes comparably long to $T$ (illustrated by the green curve in Fig.~\ref{fig:A2} flattening out compared to the purple curve starting from $N \approx 6$).

\clearpage
\section{Effect of not including SPAM errors in the likelihood model}
\begin{figure}
    \centering
    \includegraphics[width=0.9\textwidth]{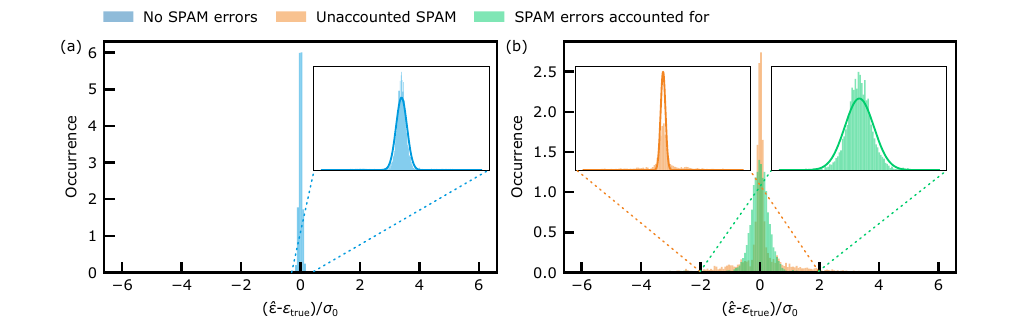}
    \caption{\textbf{Effect of not including SPAM errors in the likelihood model.} \textbf{(a)} Simulated distribution of estimation errors after $N=15$ steps without SPAM errors. \textbf{(b)} Simulated estimation errors after $N=15$ steps with SPAM errors modeled using $\alpha=-0.02$ and $\beta=0.6$, where the orange histogram shows the distribution of errors if the Bayesian update is performed assuming no SPAM errors (i.e $\alpha=0$ and $\beta=1$) while the green histogram shows the distribution of errors when the correct values $\alpha=-0.02$ and $\beta=0.6$ are used in the Bayesian update model. Insets show zoomed-in histograms together and the Gaussian distributions of errors that the algorithm expects to see given all the final posterior variances (solid lines). All histograms consist of 5,000 simulated experiments and are normalized.}
    \label{fig:FigS4_FBS}
\end{figure}

The Bayesian scheme presented in this work only focuses on estimation of a single parameter in order to obtain a simple parametric solution to the update equations, assuming reasonably good estimates for the other parameters of the model are known beforehand, by fitting to the data of simpler preliminary experiments. In the context of the FBS, this means providing estimates of the dephasing parameter $T$ and the SPAM error coefficients $\alpha$ and $\beta$. Here we investigate through simulations how neglecting SPAM errors in the model for how the measurement data is generated impacts estimation errors in the case of a model where such SPAM errors are present. Fig.~\ref{fig:FigS4_FBS}(a) shows the simulated distribution of estimation errors from the true frequency if perfect initialization and readout could be achieved, i.e., the data generating function is
\begin{equation}
\label{eq:no_spam}
	P(m|\varepsilon,\Updelta f,\tau)=\frac{1}{2}\left(1+me^{-\tau/T}\cos{(2\pi(\Updelta f-\varepsilon)\tau)}\right),
\end{equation} 
while Fig.~\ref{fig:FigS4_FBS}(b) shows simulated estimation error distributions when measurements are generated according to
\begin{equation}
\label{eq:with_spam}
	P(m|\varepsilon,\Updelta f,\tau)=\frac{1}{2}\left(1+m[\alpha+\beta e^{-\tau/T}\cos{(2\pi(\Updelta f-\varepsilon)\tau)}]\right),
\end{equation}
with $\alpha=-0.02$ and $\beta=0.6$, where the orange histogram is the result when mistakenly assuming the model Eq.~\eqref{eq:no_spam} and the green histogram is the result when updating according to the correct model in Eq.~\eqref{eq:with_spam}. We see that updating according to the wrong model parameters $\alpha$ and $\beta$ results in tails of large estimation errors. A large mismatch in the true value of $\alpha$ and its estimated value would have the effect of biasing the frequency estimation error towards positive or negative values, but here $\alpha$ is small enough to not to have much impact. Accounting for the SPAM errors by choosing the correct values of $\alpha$ and $\beta$ has the effect of reducing the occurrence of large outlier errors and any estimation error bias. The errors are still larger than in the case of no SPAM errors, which is due to the update equations becoming more conservative when reducing the posterior variance. This is a result of the information gathered from each measurement becoming less certain, and so more measurements are necessary to obtain as low estimation errors. The fact that most of the estimation errors can still be reduced by more measurements is an important contrast to the case of the simpler model that does not account for the SPAM errors present. Even though this simpler model has more occurrences of very low errors, the many outliers in the tails exceed the amount of large errors the algorithm would expect if it was correct, with these outliers presumably being estimates that most often cannot be improved much by further gathering of measurements. These differences in what sort of estimation errors the algorithm expects versus what errors are actually realized are shown in the insets of Fig.~\ref{fig:FigS4_FBS}. The simulations here consist of 5,000 attempts to get estimates $\hat{\varepsilon}$ for different true values $\varepsilon_{\text{true}}$ drawn from the initial prior, which here is the standard normal distribution, ie. $\varepsilon_{\text{true}}\sim \mathcal{N}(0,1)$.

\clearpage

\section{Validity of the Gaussian Approximation}
\begin{figure}[h]
    \centering
    \includegraphics[width=\linewidth]{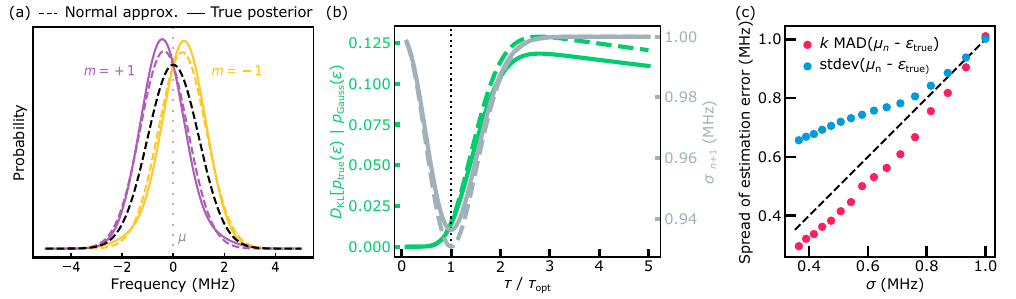}
    \caption{
    \textbf{Validity of the Gaussian approximation.}
    \textbf{(a)} Approximated Gaussian distributions (dashed lines) versus true posteriors (solid lines) from a Gaussian prior of variance $\SI{1}{\mega\hertz^2}$ (black dashed lines), with a dephasing time $T=\SI{10}{\micro\second}$ and SPAM errors $\alpha=-0.02$ and $\beta=0.6$. While the oscillatory likelihood adds some skewness and broader tails, the posteriors are close to Gaussian, with deviations mainly seen in small shifts and kurtosis mismatches. \textbf{(b)} KL divergence between the true posterior and its Gaussian method of moments fit and the posterior standard deviation $\sigma$ for the measurement outcomes $m=-1$ (dashed) and $m=+1$ (solid) at different choices of $\tau$. \textbf{(c)} Posterior standard deviation as an estimator of error. The posterior distributions in the Bayesian framework provide estimators for the error in the estimate $\varepsilon$, here given by $\sigma$. Due to a few outliers of large errors, $\sigma$ turns out to be a poor estimate for the standard deviation of errors, ie. it does not follow the diagonal black line. Instead, $\sigma$ is a good estimator of the median absolute deviation (assuming the measurement generating model is correct), which is related to $\sigma$ by a factor $k$ (for normally distributed data $k=1.4826$).}
    \label{fig:FigS5_FBS}
\end{figure}

We approximate the true posterior, starting from a Gaussian prior with an oscillating likelihood, as Gaussian. However, this assumption may not always hold, especially for certain parameter choices where the oscillatory likelihood can create posteriors with multiple peaks. For instance, specific values of $\tau$ can lead to multimodal posteriors. Here we show that an optimal choice of $\tau$ results in fairly Gaussian posteriors, as shown in Fig.~\ref{fig:FigS5_FBS}(a). To quantify how different the true posterior is from the Gaussian we use to approximate it we can calculate the Kullback–Leibler (KL) divergence
\begin{equation}
    \mathcal{D}_{\text{KL}}(p||q) = \int_{-\infty}^{\infty} p(\varepsilon) \log_{2}\left(\frac{p(\varepsilon)}{q(\varepsilon)}\right) d\varepsilon,
\end{equation}
with $p$ here denoting the true distribution and $q$ the Gaussian approximation. In Fig.~\ref{fig:FigS5_FBS}(b) we plot the KL divergence (green) for a range of values $\tau$ around the optimal value, together with the resulting posterior standard deviations (black), for the case of $m=-1$ (dashed) and $m=1$ (solid). At the value $\tau_{\text{opt}}$ that minimizes the expected posterior standard deviation the KL divergence is still relatively small, indicating that the true posterior still looks much like the approximated Gaussian. 

We select $\Updelta f$ to align the likelihood’s inflection point with the prior’s mean, and then choose $\tau$ to minimize the posterior variance. This choice makes the likelihood’s period comparable with the prior’s standard deviation, smoothing out the oscillations in the prior's tails.

To see how well the always Gaussian approximation works across an estimation run of $N=15$ single-shot measurements we plot the posterior standard deviation versus the actual spread of errors, quantified in two different ways, as seen in Fig.~\ref{fig:FigS5_FBS}(c). The $16$ dots for each of the two quantities show the error for $N=0$ to $N=15$ measurements (positioned right to left). The data here is generated by performing 5,000 simulations, where we have assumed that the likelihood function Eq.~\eqref{eq:with_spam} with $\alpha=-0.02$, $\beta=0.6$ and $T=\SI{10}{\micro\second}$ is indeed the correct measurement generating function. It turns out that $\sigma$ seems to be a poor estimator for the actual standard deviation of the errors (blue dots). This is presumably because the standard deviation is very sensitive to outliers, and these outliers (although still quite rare) occur more often than expected due to approximating the true posteriors by Gaussians which have slightly slimmer tails. However, as seen in Fig.~\ref{fig:FigS4_FBS} the error distribution still overlaps very well with a zero-centered Gaussian with the average $\sigma$ as standard deviation. The posterior standard deviation $\sigma$ instead turns out to be quite a good estimator for the median absolute deviation (MAD), which is less sensitive to outliers. One can relate the MAD to a standard deviation $\sigma$ by a factor $k$, so $\sigma = k \cdot \text{MAD}$. To be consistent with a normal distribution one sets $k=1.4826$~\cite{huber_robust_1981}. Thus the Gaussian approximation works well as long as it is acceptable to have some more outliers in terms of estimation error compared to what would be generated if the distribution truly was Gaussian (although the number of outliers is still small compared to expected outcomes).

	\nocite{*}
	
	\bibliography{suppl_my_bibliography}